\documentclass[fleqn,usenatbib]{mnras}

\usepackage{newtxtext,newtxmath}
\usepackage[T1]{fontenc}
\usepackage{ae,aecompl}
\usepackage[caption=false]{subfig}
\usepackage{tikz}
\usepackage{standalone}

\usepackage{graphicx}	% Including figure files
\usepackage{amsmath}	% Advanced maths commands
\usepackage{amssymb}	% Extra maths symbols
\usepackage{booktabs}

\def\beq{\begin{eqnarray}}
\def\eeq{\end{eqnarray}}

\newcommand{\av}[1]{\langle{#1\rangle}} 
\let\vec\mathbf

\numberwithin{equation}{section}

\title[Covariance Matrices in Arbitrary Survey Geometries]{Estimating Covariance Matrices for Two- and Three-Point Correlation Function Moments in Arbitrary Survey Geometries}

\author[O.\,H.\,E. Philcox \& D.\,J. Eisenstein]{
Oliver H.\,E. Philcox$^{1,2}$\thanks{E-mail: \href{mailto:ohep2@alumni.cam.ac.uk}{ohep2@alumni.cam.ac.uk}}
and Daniel J. Eisenstein$^{1}$
\\
$^{1}$Center for Astrophysics | Harvard \& Smithsonian, 60 Garden St., MA 02138, USA\\
$^{2}$Department of Astrophysical Sciences, Princeton University, Princeton, NJ 08544, USA
}

\date{Accepted 2019 October 9. Received 2019 September 24; in original form 2019 June 19}

\pubyear{2019}

\begin{document}
\label{firstpage}
\pagerange{\pageref{firstpage}--\pageref{lastpage}}
\maketitle

% Abstract of the paper
\begin{abstract}
We present configuration-space estimators for the auto- and cross-covariance of two- and three-point correlation functions (2PCF and 3PCF) in general survey geometries. These are derived in the Gaussian limit (setting higher-order correlation functions to zero), but for arbitrary non-linear 2PCFs (which may be estimated from the survey itself), with a shot-noise rescaling parameter included to capture non-Gaussianity. We generalize previous approaches to include Legendre moments via a geometry-correction function calibrated from measured pair and triple counts. Making use of importance sampling and random particle catalogs, we can estimate model covariances in fractions of the time required to do so with mocks, obtaining estimates with negligible sampling noise in $\sim 10$ ($\sim 100$) CPU-hours for the 2PCF (3PCF) auto-covariance. We compare results to sample covariances from a suite of BOSS DR12 mocks and find the matrices to be in good agreement, assuming a shot-noise rescaling parameter of $1.03$ ($1.20$) for the 2PCF (3PCF). To obtain strongest constraints on cosmological parameters we must use multiple statistics in concert; having robust methods to measure their covariances at low computational cost is thus of great relevance to upcoming surveys.
\end{abstract}

% Select between one and six entries from the list of approved keywords.
% Don't make up new ones.

%%%%%%%%%%%%%%%%%%%%%%%%%%%%%%%%%%%%%%%%%%%%%%%%%%

%%%%%%%%%%%%%%%%% BODY OF PAPER %%%%%%%%%%%%%%%%%%

\section{Introduction}\label{sec: intro}

In the standard cosmological paradigm, where density fluctuations are created in reheating following inflation, the primordial over-density field, $\delta$, is expected to be Gaussian and thus describable completely by its (isotropic) power spectrum, or associated two-point correlation function (2PCF), $\xi(r)$. During the later evolution of the Universe, gravitational effects introduce non-Gaussianities into $\delta$, implying that higher-order statistics are required to fully describe the field, the prime candidate being the isotropic three-point correlation function (3PCF) (see \citet{2002PhR...367....1B} and \citet{2005astro.ph..5391S} for reviews). There is a wealth of research concerning both the expected forms of the configuration-space 3PCF \citep[e.g.][]{2002PhR...367....1B,2006PhRvD..74b3522S,2008ApJ...672..849M,2017MNRAS.469.2059S,2017MNRAS.472..577Y,2018MNRAS.478.2019Y} and estimators for its measurement \citep[e.g.][]{1998ApJ...494L..41S,2001misk.conf...71M,2004ASPC..314..249G,2004ApJ...605L..89S,2006MNRAS.368.1507N,2015MNRAS.454.4142S}, and it has been shown that this observable is able to break degeneracies between cosmic parameters \citep[e.g.][]{1994ApJ...437L..13G,1994ApJ...425..392F,2004ApJ...607..140J,2015MNRAS.449L..95G} and provide probes of primordial non-Gaussianity \citep{2010CQGra..27l4011D}. Because of this, the isotropic 3PCF has become a powerful addition to the analyst's toolkit, and has been used in a number of analyses \citep[e.g.][]{2004PASJ...56..415K,2004ApJ...607..140J,2011ApJ...737...97M,2015arXiv151202231S,2017MNRAS.469.1738S,2018MNRAS.474.2109S}.

Despite the Universe's statistical isotropy, the measured 2PCF and 3PCF are far from isotropic, both due to Redshift Space Distortions (RSD), caused by degeneracies between cosmic and peculiar velocity redshifts, and the Alcock-Paczynski effect, where anisotropy is introduced due to incorrectly assumed cosmology \citep{1979Natur.281..358A}. In particular, the Finger-of-God (Kaiser) effect boosts (reduces) the power parallel to the line of sight (LoS), resulting in a LoS-dependent correlation function \citep{1972MNRAS.156P...1J,1987MNRAS.227....1K}. Na\"{i}vely, this may seem to be simply an irritation, yet in practice anisotropy provides an independent probe of structure formation, tightening the constraints on cosmological parameters. Due to RSD, the 2PCF becomes a two-dimensional function $\xi(r,\mu)$, where $\mu$ is the angle of a pair of galaxies to the LoS, whilst the 3PCF now depends on an additional two parameters describing the orientation of the triangle of galaxies to the LoS \citep{1999ApJ...527....1S}. Due to the high dimensionality of the anisotropic 3PCF, its measurements are inherently noisy, thus, despite there existing a multitude of algorithms for its estimation \citep[e.g.][]{2005NewA...10..569Z,2007arXiv0709.1967G,2018MNRAS.478.1468S}, it has been seldom used in cosmological analyses thus far. In contrast, the anisotropic 2PCF is a widely adopted quantity, commonly presented in one of two forms; (a) in $(r,\mu)$-space wedges (recent examples being \citealt{2012MNRAS.419.3223K} and \citealt{2017MNRAS.464.1640S}) or (b) decomposed into Legendre moments $\xi_\ell(r)$ \citep[recently][]{2017MNRAS.469.1369S}. The latter form is particularly useful since most information is captured by the first few (even) multipoles, obviating the need for a large number of (noisy) angular bins.

To constrain cosmological parameters, we must compare the above statistics to established models, requiring good understanding of the covariance matrices of our measurements. Furthermore, stronger parameter bounds are obtained by using multiple statistics (e.g. both the 2PCF and 3PCF), with optimal analyses requiring full knowledge of the \textit{cross-covariance} matrices. Any such covariance is relatively easy to derive for uniform periodic simulation boxes (e.g. see \citet{2019arXiv190806234S} for perturbation theory covariances in Fourier space), yet the selection functions of realistic surveys are far from simple, with power on a range of scales due to effects such as non-uniform boundaries and fiber collisions. The survey geometry has a significant impact on the output covariances, thus ignoring it will lead to distortions in parameter error bars. To account for this, covariance matrices are usually computed using a set of `mock' catalogs created with the same selection functions as the observational data. Creating high-fidelity mock catalogs is computationally expensive, especially since we require a large number of mocks to avoid noise in the covariance matrices biasing the derived parameter error bars \citep{2013MNRAS.432.1928T,2013AJ....145...10D,2014MNRAS.439.2531P}. In addition, future surveys (e.g. Euclid \citep{2011arXiv1110.3193L}, DESI \citep{2013arXiv1308.0847L} and WFIRST \citep{2015arXiv150303757S}) will perform tomographic analyses, greatly increasing the number of correlation function bins, thus requiring far more mocks for the same level of covariance matrix noise. Approximate methods for covariance generation can partially alleviate this problem; these permit fast estimation of low-noise matrices, allowing us to create fewer mocks of higher quality, which remain important for analysis of systematics.  

In \citet{2016MNRAS.462.2681O}, it was shown that the 2PCF covariance matrix for an arbitrary survey geometry could be written as a sum of configuration-space integrals (similar to \citealt{1994ApJ...424..569B}), and quickly evaluated using importance sampling techniques. Although the basic formalism assumes Gaussianity, good agreement of theoretical and mock-based covariances was found using a few percent rescaling of the shot-noise amplitude, calibrated either from a small suite of mocks, or from the survey itself via jackknifes \citep{2019MNRAS.487.2701O}. The formalism was further developed in \citet{rascalC}, where a new algorithm was introduced, \texttt{RascalC},\footnote{This has been extended to include the Legendre-binned 2PCF and 3PCF auto-covariances discussed in this paper. Full documentation is provided at \href{http://RascalC.readthedocs.io}{RascalC.readthedocs.io}.} allowing for single- and multi-tracer 2PCF covariances to be computed from a single survey in a fraction of the previous computation time. Our goal in this work is to extend the previous formalism to more complex cases; the auto- and cross-covariance matrices of anisotropic 2PCF and isotropic 3PCF Legendre moments. We introduce a geometry-correction function to allow for moments to be computed directly (rather than in post-processing) and demonstrate that our output precision matrices are in good agreement with those from simulations. Via efficient importance sampling methods we are able to produce 2PCF (3PCF) auto-covariance matrices with negligible sampling noise in $\sim 10$ ($\sim 100$) CPU-hours.

This paper is structured as follows. We begin with an overview of previous 2PCF estimators and covariances in Sec.\,\ref{sec: previous-work}, before discussing the extension to the covariance of the Legendre moments of single- and multi-field 2PCFs in Sec.\,\ref{sec: 2PCF-leg}, along with validation using BOSS DR12 simulations. Sec.\,\ref{sec: 3PCF} describes the 3PCF estimator adopted herein, before its theoretical covariance, and cross-covariance with the 2PCF, is considered in Sec.\,\ref{sec: 3PCF-covs}, along with application to BOSS DR12 mocks. We conclude with a summary and discussion in Sec.\,\ref{sec: Conclusion}, with a useful theoretical result presented in appendix \ref{sec: 3PCF-cancellation}.

\section{Overview of Previous Work}\label{sec: previous-work}
We begin by summarizing the estimators for the 2PCF and the formalism for the covariance matrices, as in \citet{2016MNRAS.462.2681O}, \citet{2019MNRAS.487.2701O} and \citet{rascalC}, with slightly modified notation. For a galaxy survey measuring a total of $N_\mathrm{gal}$ galaxies with continuous (mean) number density and weight functions $n(\vec r)$ and $w(\vec r)$, we may define the anisotropic 2PCF as a ratio of pair counts
\beq\label{eq: Landy-Szalay}
    \hat\xi(r,\mu) = \frac{NN(r,\mu)}{RR(r,\mu)}
\eeq
\citep{1993ApJ...412...64L}, where $\mu = \cos\theta$ measures the angular coordinate (using symmetry to restrict to $\mu\in[0,1]$) and $N = (D - R)$, with $D$ and $R$ representing galaxies and random particles respectively. In some radial bin $a$ and angular bin $c$, the 2PCF estimate becomes
\beq\label{eq: Landy-Szalay-binned}
    \hat\xi^a_c = \frac{NN^a_c}{RR^a_c}
\eeq
where $NN^a_c$ and $RR^a_c$ are defined by
\beq\label{eq: NN-RR-discrete}
    NN^a_c &=& \sum_{i\neq j}n_in_jw_iw_j\Theta^a(r_{ij})\Theta^c(\mu_{ij})\delta_i\delta_j\\\nonumber
    RR^a_c &=& \sum_{i\neq j}n_in_jw_iw_j\Theta^a(r_{ij})\Theta^c(\mu_{ij}),
\eeq
(for $X_i\equiv X(\vec r_i)$), dividing the survey into small cubic boxes which contain at most one galaxy and summing. Here $\delta_i$ is the fractional galaxy overdensity in cell $i$, $\mu_{ij}$ is the angle between the vector $\vec r_i - \vec r_j$ and the LoS (or the $z$-axis for a cuboidal simulation), and the binning functions $\Theta^a(Y)$ are unity if $Y$ is in the bin $a$ and zero else. Unlike in previous works, we expand the binning function into a radial and angular part, for later convenience. Equivalently, the pair counts may be written in continuous form;
\beq\label{eq: NN-RR-continuous}
    NN^a_c &=& \int d^3\vec r_i d^3 \vec r_j\,n(\vec r_i)n(\vec r_j)w(\vec r_i)w(\vec r_j)\Theta^a(|\vec r_i-\vec r_j|)\Theta^c(\vec \mu_{(\vec r_i-\vec r_j)})\delta(\vec r_i)\delta(\vec r_j)\\\nonumber
    RR^a_c &=& \int d^3\vec r_i d^3 \vec r_j\,n(\vec r_i)n(\vec r_j)w(\vec r_i)w(\vec r_j)\Theta^a(|\vec r_i-\vec r_j|)\Theta^c(\vec \mu_{(\vec r_i-\vec r_j)}).
\eeq

Given the 2PCF estimator, we define the covariance matrix $C^{ab}_{cd}$ in radial bins $(a,b)$ and $\mu$-bins $(c,d)$ as 
\beq\label{eq: cov2x2 defn}
    \operatorname{cov}(\hat\xi^a_c\,,\,\hat\xi^b_d) &\equiv& C^{ab}_{cd} = \av{\hat\xi^a_c\hat\xi^b_d}-\av{\hat\xi^a_c}\av{\hat\xi^b_d}\\\nonumber
    &=& \frac{1}{RR^a_cRR^b_d}\sum_{i\neq j}\sum_{k\neq l}n_in_jn_kn_lw_iw_jw_kw_l\Theta^a(r_{ij})\Theta^c(\mu_{ij})\Theta^b(r_{kl})\Theta^d(\mu_{kl})\left[\av{\delta_i\delta_j\delta_k\delta_l}-\av{\delta_i\delta_j}\av{\delta_k\delta_l}\right],
\eeq
inserting the expanded 2PCF estimator. In \cite{2016MNRAS.462.2681O}, it was shown that this could be expanded into 2-, 3- and 4-point summations using the relation
\beq\label{eq: cov2x2 expansion}
    \sum_{i\neq j}\sum_{k\neq l}X_{ij}Y_{kl} = \sum_{i\neq j\neq k\neq l}X_{ij}Y_{kl} + 4\sum_{i\neq j\neq k}X_{ij}Y_{kj} + 2\sum_{i\neq j}X_{ij}Y_{ij}
\eeq
for summands $X_{ij}$ and $Y_{kl}$ symmetric under $i\leftrightarrow j$ and $k\leftrightarrow l$ interchanges. To expand out the expectation terms, we apply the shot-noise approximation
\beq\label{eq: shot-noise}
    \delta_i^2 \approx \frac{\alpha}{n_i}(1+\delta_i)
\eeq
(including a shot-noise rescaling factor $\alpha$ to capture non-Gaussianity as discussed below) and Isserlis' (Wick's) theorem \citep{isserlis} giving
\beq\label{eq: Wicks-2x2 expansions}
    \av{\delta_i\delta_j\delta_k\delta_l}-\av{\delta_i\delta_j}\av{\delta_k\delta_l} &=& \xi^{(4)}_{ijkl}+\xi_{ik}\xi_{jl}+\xi_{il}\xi_{jk} \\\nonumber
    \av{\delta_i\delta_j\delta_k\delta_j} &\approx& \frac{\alpha}{n_j}\av{(1+\delta_j)\delta_i\delta_k} = \frac{\alpha}{n_i}\left(\zeta_{ijk}+\xi_{ik}\right)\\\nonumber
    \av{\delta_i\delta_j\delta_i\delta_j} &\approx& \frac{\alpha^2}{n_in_j}\av{(1+\delta_i)\delta_j\delta_k} = \frac{\alpha^2}{n_in_j}\left(1+\xi_{ij}\right),
\eeq
where $\zeta$ and $\xi^{(4)}$ are the three- and four-point connected correlation functions. Inserting Eqs.\,\ref{eq: cov2x2 expansion}\,\&\,\ref{eq: Wicks-2x2 expansions} into Eq.\,\ref{eq: cov2x2 defn} gives the full covariance
\beq\label{eq: Cov2x2estimator}
    C^{ab}_{cd} = {}^4C^{ab}_{cd} + \alpha \times {}^3C^{ab}_{cd} + \alpha^2 \times {}^2C^{ab}_{cd}
\eeq
with the definitions
\beq\label{eq: Cov2x2 234 Point Defs}
    {}^4C^{ab}_{cd} &=& \frac{1}{RR^a_cRR^b_d}\sum_{i\neq j\neq k\neq l}n_in_jn_kn_lw_iw_jw_kw_l\Theta^a(r_{ij})\Theta^c(\mu_{ij})\Theta^b(r_{kl})\Theta^d(\mu_{kl})\left[\xi^{(4)}_{ijkl}+2\xi_{ik}\xi_{jl}\right]\\\nonumber
    {}^3C^{ab}_{cd} &=& \frac{4}{RR^a_cRR^b_d}\sum_{i\neq j\neq k}n_in_jn_kw_i\left(w_j\right)^2w_k\Theta^a(r_{ij})\Theta^c(\mu_{ij})\Theta^b(r_{jk})\Theta^d(\mu_{jk})\left[\zeta_{ijk}+\xi_{ik}\right]\\\nonumber
    {}^2C^{ab}_{cd} &=& \frac{2\delta^{ab}\delta^{cd}}{RR^a_cRR^b_d}\sum_{i\neq j}n_in_j\left(w_iw_j\right)^2\Theta^a(r_{ij})\Theta^c(\mu_{ij})\left[1+\xi_{ij}\right]
\eeq
(cf.\,\citealt[Eqs.\,2.15-2.18]{2016MNRAS.462.2681O}), where $\delta^{ab}$ and $\delta^{cd}$ are Kronecker deltas. As in \citet{rascalC}, we have replaced $\xi_{ik}\xi_{jl}+\xi_{il}\xi_{jk}$ with $2\xi_{ik}\xi_{jl}$ in the 4-point integral, exploiting the relabelling symmetries of the summand. In previous work, we take the Gaussian limit, setting all higher-order correlation functions to zero, such that only the Gaussian (yet non-linear) 2PCF $\xi_{ij}$ is used in the above expressions, with shot-noise rescaling (i.e. $\alpha>1$) used to encapsulate the higher-point effects. In the expressions above, we include all terms for clarity, and note that these may be translated into continuous form by replacing $\sum_i\rightarrow \int d^3\vec r_i$ and promoting quantities to be continuous functions of the spatial position $\vec r$ e.g. $n_i\rightarrow n(\vec r_i)$ and $\xi_{ij}\rightarrow \xi(\vec r_i -\vec r_j)$. 

Before extending these techniques further, it is worth pausing to consider our main assumption, that non-Gaussianity can be well approximated using a rescaling of galactic shot-noise. This is motivated by noting that the principal action of the higher-point correlation function terms (that source non-Gaussianity) is to provide additional clustering power at small distances, usually less than the covariance binning width. By artificially increasing the level of shot-noise, we boost the clustering on \textit{infinitesimally} small scales, which is found to be a fair approximation in practice. The non-Gaussian covariance matrix terms depend on integrals of the higher-point correlation functions; our assumption is equivalent to replacing these with their contractions, giving terms identical in form to those from shot-noise. Mathematically, this is a good approximation if we assume the correlation functions to be dominated by their squeezed limits on the relevant scales. Whether shot-noise rescaling adequately describes non-Gaussianity is uncertain \textit{a priori}, but previous works \citep{2016MNRAS.462.2681O,2019MNRAS.487.2701O,rascalC} have demonstrated excellent agreement between the theoretical covariances and those of large suites of mocks, as well as the shown the method's utility in BAO-scale analyses \citep{2018MNRAS.477.1153V}.

\section{Two Point Function Covariances in Legendre Polynomial Bins}\label{sec: 2PCF-leg}
\subsection{Legendre Moment Estimator for the 2PCF}\label{subsec: 2PCF-Legendre}
In previous works the anisotropic 2PCF has been assumed to be measured as a function of the angular coordinate $\mu$, yet in many analyses \citep[e.g.][]{2001Natur.410..169P,2016arXiv161100036D,2017MNRAS.470.2617A,2017MNRAS.464.1640S,2018MNRAS.477.1639Z} it is presented instead in terms of its Legendre moments $\xi_\ell(r)$, which can be simply related to the power spectrum multipoles. Conversion between $\xi_\ell(r)$ and $\xi(r,\mu)$ is achieved via the standard relation
\beq
    \xi(r,\mu) = \sum_{\ell=0}^{\infty} \xi_\ell(r)L_\ell(\mu)
\eeq
where $L_\ell(\mu)$ is the Legendre polynomial of order $\ell$. Although this provides a way to convert from angularly-binned 2PCF estimates to Legendre moments, it is much more efficient to compute the latter quantities directly, avoiding errors from finite $\mu$ binning and the necessity for a finely binned $\xi(r,\mu)$. Via completeness of the Legendre polynomials (i.e. $\int_{-1}^1d\mu L_\ell(\mu)L_{\ell'}(\mu) = 2\delta_{\ell\ell'}/(2\ell+1)$), the above relation may be inverted to yield
\beq\label{eq: Legendre-from-angular-2PCF}
    \xi_\ell(r) &=& \frac{2\ell+1}{2}\int_{-1}^1d\mu\, \xi(r,\mu)L_\ell(\mu) =\frac{(-1)^\ell+1}{2}(2\ell+1)\int_0^1d\mu\, \xi(r,\mu)L_\ell(\mu)
\eeq
using $\xi(r,\mu) = \xi(r,-\mu)$ and the symmetry properties of $L_\ell(\mu)$. From the prefactor, we note that all odd $\ell$ terms vanish, thus we restrict to even $\ell$ henceforth, using a total of $N_\ell = \ell_\mathrm{max}/2+1$ bins. Restricting to a radial bin $a$ (with $r\in[r_{a,\mathrm{min}},r_{a,\mathrm{max}}]$ and center $r_a$) this becomes
\beq\label{eq: Legendre-from-angular-2PCF-v2}
    \xi_\ell^a = (2\ell+1)\int_0^1d\mu\,\xi^a(\mu)L_\ell(\mu).
\eeq
where the function $\xi^a(\mu)$, binned only in the radial coordinate, is defined by the reduced Landy-Szalay estimator
\beq
    \hat\xi^a(\mu) = \frac{NN^a(\mu)}{RR^a(\mu)}
\eeq
as before (Eqs.\,\ref{eq: Landy-Szalay}\,\&\,\ref{eq: Landy-Szalay-binned}). The quantities $XX^a(\mu)$ (for $X\in\{D,R\}$) may be identified with pair-counts in vanishingly small angular bins $\{c\}$ via
\beq\label{eq: continuous-binned counts}
    XX^a_c = \int d\mu\,XX^a(\mu)\Theta^c(\mu) \approx XX^a(\mu_c)\delta \mu
\eeq
where $\mu_c$ is the center of bin $c$ with width $\delta\mu$. This becomes exact in the limit $\delta\mu\rightarrow0$. This may be inverted to yield
\beq\label{eq: binned-continuous counts}
    XX^a(\mu) \approx \sum_c\frac{XX^a_c}{\delta\mu}\Theta^c(\mu)
\eeq

To proceed, we require an analytic form for the $RR$ pair counts. For a uniform infinite survey, the continuous form of $RR^a_c$ (Eq.\,\ref{eq: NN-RR-continuous}) may be rewritten as
\beq
    RR^a_c = (nw)^2\int d^3\vec r_id^3\vec r_j\,\Theta^a(|\vec r_i - \vec r_j|)\Theta^c(\mu_{(\vec r_i - \vec r_j)})
\eeq
since $n$ and $w$ will be independent of position. Substituting $\vec x = \vec r_i - \vec r_j$ gives
\beq
    RR^a_c &=& (nw)^2\left[\int d^3\vec r_j\right]\left[\int d^3\vec{x} \Theta^a(|\vec x|)\Theta^c(\mu_{x})\right]\\\nonumber
    &=& (nw)^2 \times V \times \int x^2dx\,\Theta^a(x) \times \int_{0}^{2\pi} d\phi_x \times \int_{-1}^1d\mu_x\Theta^c(\mu_x)\\\nonumber
    &\approx& (nw)^2\times V \times \frac{1}{3}\left(r_{a,\mathrm{max}}^3-r_{a,\mathrm{min}}^3\right) \times 2\pi \times 2\delta\mu
\eeq
defining $V$ as the survey volume and noting that an additional factor of $2$ arises in the $\mu$ binning, since we do not distinguish between positive and negative $\mu$. Similar to \citet{2010ApJ...718.1224X}, we introduce a \textit{survey correction factor} $\Phi(r_a,\mu)$ for each radial bin $a$ to encapsulate the effects of non-uniform sampling and the survey boundaries, defining
\beq\label{eq: RR_approximation}
    RR^a(\mu) \equiv \frac{\frac{4\pi}{3} V\overline{(nw)^2}\left(r_{a,\mathrm{max}}^3-r_{a,\mathrm{min}}^3\right)}{\Phi(r_a,\mu)}
\eeq
(converting to continuous form in $\mu$ via Eq.\,\ref{eq: continuous-binned counts}), where $\overline{(nw)^2}$ is the survey-averaged value of $(nw)^2$. In \citet{2010ApJ...718.1224X}, the associated function $\Phi(r,\mu)$ (continuous in both $r$ and $\mu$) was shown to be well approximated by a smooth two-dimensional function; here we may use a distinct function for each radial bin, computed by fitting the ratio of our infinite model and empirically derived pair counts, e.g. from \texttt{corrfunc}\footnote{\href{http://corrfunc.readthedocs.io}{corrfunc.readthedocs.io}} \citep{2017ascl.soft03003S}. In the case of an infinite survey we expect $\Phi(r_a,\mu) = 1$ for all $(r_a,\mu)$, though in real surveys stronger departures from $\Phi = 1$ are expected for larger radial bins, due to more pairs coming close to the survey boundaries. $\Phi$ is hence a useful function, since it allows us to simply assess the impact of the survey window on the pair counts at different radial separations and angles, and effectively `convert' between windowed and ideal surveys.

We proceed to insert this into the Legendre-binned 2PCF estimator;
\beq
    \hat\xi^a_\ell = \frac{2\ell+1}{V\overline{(nw)^2}v_a}\int_0^1 d\mu L_\ell(\mu) \
    \sum_c\Theta^c(\mu)\Phi(r_a,\mu)\frac{NN^a_c}{\delta \mu}
\eeq
converting $NN^a(\mu)$ to its binned form via Eq.\,\ref{eq: binned-continuous counts} and denoting the shell volume as $v_a = \tfrac{4\pi}{3}\left(r^3_{a,\mathrm{max}}-r^3_{a,\mathrm{min}}\right)$. The integral can be converted to an additional sum over $\mu$ bins of (arbitrarily small) width $\delta\mu$ with centers at $\mu_n$;
\beq
    \hat\xi^a_\ell = \frac{2\ell+1}{V\overline{(nw)^2}v_a}\sum_n \delta\mu\,L_\ell(\mu_n) \sum_c \Theta^c(\mu_n)\Phi(r_a,\mu_n)\frac{NN^a_c}{\delta\mu}.
\eeq
We now insert the $NN^a_c$ estimator (Eq.\,\ref{eq: NN-RR-discrete}) which yields
\beq
    \hat\xi^a_\ell = \frac{2\ell+1}{V\overline{(nw)^2}v_a}\sum_n\sum_c\sum_{i\neq j}\Theta^c(\mu_n)\Theta^c(\mu_{ij})\Theta^a(r_{ij})L_\ell(\mu_n)\Phi(r_a,\mu_n)n_in_jw_iw_j\delta_i\delta_j.
\eeq
Taking the limit $\delta\mu\rightarrow0$, the binning functions tend to Dirac delta functions which allows us to identify $\mu_n = \mu_{ij}$, giving the simplified Legendre-binned 2PCF estimator;
\beq\label{eq: 2PCF in Legendre Moments}
    \hat\xi_\ell^a = \frac{2\ell+1}{V\overline{(nw)^2}v_a}\sum_{i\neq j}n_in_jw_iw_j\Theta^{a}(r_{ij})\Phi(r_a,\mu_{ij})L_\ell(\mu_{ij})\delta_i\delta_j
\eeq
(valid for even $\ell$). Notably, the survey correction factor $\Phi$ depends on the radial bin center $r_a$ rather than the radial separation of each pair of points, $r_{ij}$, since it arises from the $RR^a(\mu)$ pair counts. Although we have only integrated over $\mu$ here, this formalism can simply be extended to compute similar estimators for $r$-integrated statistics, such as the Baryon Acoustic Oscillation (BAO) $\omega_\ell$ statistic introduced in \citet{2010ApJ...718.1224X}.

\subsection{2PCF Covariance Matrix}\label{subsec: 2pcf_covariance}
Using the Legendre-moment 2PCF estimator above (Eq.\,\ref{eq: 2PCF in Legendre Moments}), we may define the associated covariance matrix in radial bins $a,b$ and Legendre indices $p,q$ in the standard fashion;
\beq\label{eq: legendre_cov2x2_defn}
    \operatorname{cov}(\hat\xi^a_p\,,\,\hat\xi^b_q) \equiv C^{ab}_{pq} = \av{\hat\xi^a_p\hat\xi^b_q}-\av{\hat\xi^a_p}\av{\hat\xi^b_q}.
\eeq
Using the Legendre polynomial expansions (Eq.\,\ref{eq: Legendre-from-angular-2PCF}) this can be related to the covariance in angular bins $m,n$ (i.e. Eq.\,\ref{eq: Cov2x2estimator}) via 
\beq
C^{ab}_{pq} \approx (2p+1)(2q+1)\int_0^1 d\mu \int_0^1 d\mu'\, L_p(\mu)L_q(\mu')\sum_{m,n}\Theta^{m}(\mu)\Theta^n(\mu')\left(\delta\mu\right)^{-2}C^{ab}_{mn}
\eeq
which becomes exact in the limit of infinitely small $\mu$ bins. As before, we opt to compute this directly, rather than from the angularly binned matrices. Inserting the 2PCF estimators (Eq.\,\ref{eq: 2PCF in Legendre Moments}) into Eq.\,\ref{eq: legendre_cov2x2_defn}, we obtain the estimator
\beq\label{eq: cov2x2Legendre Unsimplified}
    \hat C^{ab}_{pq} &=& \frac{(2p+1)(2q+1)}{(V\overline{(nw)^2})^2v_av_b}\sum_{i\neq j}\sum_{k\neq l}n_in_jn_kn_lw_iw_jw_kw_l\Theta^a(r_{ij})\Theta^b(r_{kl})\\\nonumber
    &\times&\Phi(r_a,\mu_{ij})\Phi(r_b,\mu_{kl})L_p(\mu_{ij})L_q(\mu_{kl})\left[\av{\delta_i\delta_j\delta_k\delta_l}-\av{\delta_i\delta_j}\av{\delta_k\delta_l}\right].
\eeq
The expansion of the above summation into 2-, 3- and 4-point terms and the application of Wick's theorem is performed exactly as for the $(r,\mu)$ binned 2PCF covariance (Eqs.\,\ref{eq: cov2x2 expansion}-\ref{eq: Wicks-2x2 expansions}), which gives the final form 
\beq\label{eq: cov2x2 Legendre estimator}
    C^{ab}_{pq} = {}^4C^{ab}_{pq} + \alpha \times {}^3C^{ab}_{pq} + \alpha^2 \times {}^2C^{ab}_{pq}
\eeq
with the definitions
\beq\label{eq: cov2x2 Legendre terms}
    {}^4C^{ab}_{pq} &=& \frac{(2p+1)(2q+1)}{(V\overline{(nw)^2})^2v_av_b}\sum_{i\neq j\neq k \neq l}n_in_jn_kn_lw_iw_jw_kw_l\Theta^a(r_{ij})\Theta^b(r_{kl})\\\nonumber
    &\times&\Phi(r_a,\mu_{ij})\Phi(r_b,\mu_{kl})L_p(\mu_{ij})L_q(\mu_{kl})\left[\xi^{(4)}_{ijkl}+2\xi_{ik}\xi_{jl}\right]\\\nonumber
    {}^3C^{ab}_{pq} &=& 4\times \frac{(2p+1)(2q+1)}{(V\overline{(nw)^2})^2v_av_b}\sum_{i\neq j\neq k}n_in_jn_kw_i\left(w_j\right)^2w_k\Theta^a(r_{ij})\Theta^b(r_{jk})\\\nonumber
    &\times&\Phi(r_a,\mu_{ij})\Phi(r_b,\mu_{jk})L_p(\mu_{ij})L_q(\mu_{jk})\left[\zeta_{ijk}+\xi_{ik}\right]\\\nonumber
    {}^2C^{ab}_{pq} &=& 2\delta_{ab}\times \frac{(2p+1)(2q+1)}{(V\overline{(nw)^2})^2v_a^2}\sum_{i\neq j}n_in_j\left(w_iw_j\right)^2\Theta^a(r_{ij})\\\nonumber
    &\times&\left(\Phi(r_a,\mu_{ij})\right)^2L_p(\mu_{ij})L_q(\mu_{ij})\left[1+\xi_{ij}\right].
\eeq
As before, these may be transformed into integrals by replacing the summations with integrals and promoting variables to be functions of position. Notably the 2-point function is no-longer diagonal, with off-diagonal elements arising for $m\neq n$, due to the Legendre binning.

Practically, the integrals in Eqs.\,\ref{eq: cov2x2 Legendre terms} may be computed using a similar algorithm to the $(r,\mu)$-binning algorithm described in \cite{rascalC}. As before, sets of four particles can be drawn according to some importance sampling scheme chosen to give efficient integral sampling. Instead of adding the integral contributions of these particles (in some radial bins $a,b$) to a single set of $\mu$ bins, we may compute contributions to \textit{all} $(p,q)$-bins simultaneously, by evaluating the relevant Legendre polynomials. This, coupled with the fact that less Legendre bins are usually required than $\mu$-bins, will significantly boost computational speed, sampling $N_\ell^2$ more bins for the same number of quads drawn, with little additional computation required.  It is important to note that this formalism does not simply extend to the jackknife covariance matrices discussed in \citet{2019MNRAS.487.2701O} and \citet{rascalC}. Since these depend on the 2PCFs computed using specific jackknive regions, we would need to introduce survey-correction functions for each individual jackknife as well as each radial bin, greatly increasing the complexity. We note however that the principal use of the jackknife matrices is in determining the shot-noise rescaling parameter $\alpha$, which is the same for Legendre-binned and $\mu$-binned full covariance matrices. We may thus use $\mu$-binned jackknife matrices to constrain $\alpha$, which can then be used to compute full Legendre-binned covariances.

\subsection{Multi-Field 2PCF Covariances}\label{subsec: 2pcf-covariance-multi}
We should additionally consider the Legendre-binned 2PCF covariance matrix arising from two different fields (cf.\,\citealt[Sec.\,6]{rascalC}). We first require an approximation for the $RR$ pair counts between the two fields, labelled $X$ and $Y$. By analogy with the above, the $R^{X}R^{Y}$ integral in radial bin $a$ and angular bin $c$ is given by 
\beq\label{eq: generalized RR_counts}
\left(R^{X}R^{Y}\right)^a_c &=& \int d^3\vec r_i\,d^3 \vec r_j\,n^X(\vec r_i)n^Y(\vec r_j)w^X(\vec r_i)w^Y(\vec r_j)\Theta^a(|\vec r_i-\vec r_j|)\Theta^c(\vec \mu_{(\vec r_i-\vec r_j)})
\eeq
where the superscripts indicate what field the quantity belongs to. As before, we use infinite uniform survey assumptions and a correction factor to yield the continuous random count form
\beq\label{eq: RR_approximation_generalized}
\left(R^{X}R^{Y}\right)^a(\mu) \equiv \frac{V^X\overline{n^Xn^Yw^Xw^Y}v_a}{\Phi^{XY}(r_a,\mu)}.
\eeq
Here $V^X$ is the volume of survey $X$, and $\overline{n^Xn^Yw^Xw^Y}$ is the mean of $n^Xn^Yw^Xw^Y$ acrsoss the survey. Notably, this appears asymmetric with respect to $X\leftrightarrow Y$ interchanges, due to presence of only a single factor of volume. The differing volumes of the two surveys are encapsulated by the correction factor $\Phi^{XY}$ and we here note that $\Phi^{XY}/V^X = \Phi^{YX}/V^Y$, restoring the field-interchange symmetry of $R^XR^Y$. The survey-correction factors $\Phi^{XY}$ can be found from comparing the $R^XR^Y$ model to computed pair counts, e.g. with \texttt{corrfunc}. From this, and adopting the Landy-Szalay generalization
\beq
    \left(\xi^{XY}\right)^a(\mu) = \frac{\left(N^XN^Y\right)^a(\mu)}{\left(R^XR^Y\right)^a(\mu)}
\eeq
where $N^T\equiv D^T-R^T$ for $T\in\{X,Y\}$, the Legendre-binned 2PCF estimator becomes
\beq\label{eq: generalized 2PCF in Legendre Moments}
    \left(\hat\xi^{XY}\right)_\ell^a = \frac{2\ell+1}{V^X \overline{n^Xn^Yw^Xw^Y}v_a}\sum_{i\neq j}n^X_in^Y_jw^X_iw^Y_j\Theta^{a}(r_{ij})\Phi^{XY}(r_a,\mu_{ij})L_\ell(\mu_{ij})\delta^X_i\delta^Y_j
\eeq
where $\delta^X$ and $\delta^Y$ are the overdensities in fields $X$ and $Y$. The full Legendre-binned covariance matrix follows similarly from this, taking the form
\beq\label{eq: theoretical_generalized_jackknife_matrix}
    \left(C^{XY,ZW}\right)^{ab}_{pq} &=& \left({}^4C^{XY,ZW}\right)^{ab}_{pq} + \frac{\alpha^X}{4} \left[\delta^{XW}\left({}^3C^{X,YZ}\right)^{ab}_{pq}+\delta^{XZ}\left({}^3C^{X,YW}\right)^{ab}_{pq}\right]+\frac{\alpha^Y}{4}\left[\delta^{YW}\left({}^3C^{Y,XZ}\right)^{ab}_{pq}+\delta^{YZ}\left({}^3C^{Y,XW}\right)^{ab}_{pq}\right]\\\nonumber
    &+&\frac{\alpha^X\alpha^Y}{2}\left(\delta^{XW}\delta^{YZ}+\delta^{XZ}\delta^{YW}\right)\left({}^2C^{XY}\right)^{ab}_{pq}.
\eeq
where $\alpha^X$ and $\alpha^{XY}$ are shot-noise rescaling parameters for fields $X$ and $Y$ and $\delta^{XY}$ etc. are Kronecker deltas. The 2-, 3- and 4-point components are defined by
\beq\label{eq: cov2x2 Legendre terms_generalized}
    \left({}^4C^{XY,ZW}\right)^{ab}_{pq} &=& \frac{(2p+1)(2q+1)}{V^XV^Z\overline{n^Xn^Yw^Xw^Y}\,\overline{n^Zn^Ww^Zw^W}v_av_b}\sum_{i\neq j\neq k \neq l}n^X_in^Y_jn^Z_kn^W_lw^X_iw^Y_jw^Z_kw^W_l\Theta^a(r_{ij})\Theta^b(r_{kl})\\\nonumber
    &\times&\Phi^{XY}(r_a,\mu_{ij})\Phi^{ZW}(r_b,\mu_{kl})L_p(\mu_{ij})L_q(\mu_{kl})\left[\xi^{(4),XYZW}_{ijkl}+\xi^{XZ}_{ik}\xi^{YW}_{jl}+\xi^{XW}_{il}\xi^{YZ}_{jk}\right]\\\nonumber
    \left({}^3C^{Y,XZ}\right)^{ab}_{pq} &=& 4\times \frac{(2p+1)(2q+1)}{V^XV^Z\overline{n^Xn^Yw^Xw^Y}\,\overline{n^Yn^Zw^Yw^Z}v_av_b}\sum_{i\neq j\neq k}n^X_in^Y_jn^Z_kw^X_i\left(w^Y_j\right)^2w^Z_k\Theta^a(r_{ij})\Theta^b(r_{jk})\\\nonumber
    &\times&\Phi^{XY}(r_a,\mu_{ij})\Phi^{ZY}(r_b,\mu_{jk})L_p(\mu_{ij})L_q(\mu_{jk})\left[\zeta^{XYZ}_{ijk}+\xi^{XZ}_{ik}\right]\\\nonumber
    \left({}^2C^{XY}\right)^{ab}_{pq} &=& 2\delta_{ab}\times \frac{(2p+1)(2q+1)}{\left(V^X\overline{n^Xn^Yw^Xw^Y}v_a\right)^2}\sum_{i\neq j}n^X_in^Y_j\left(w^X_iw^Y_j\right)^2\Theta^a(r_{ij})\\\nonumber
    &\times&\left(\Phi^{XY}(r_a,\mu_{kl})\right)^2L_p(\mu_{ij})L_q(\mu_{ij})\left[1+\xi^{XY}_{ij}\right].
\eeq
These expressions reduce to the single-field 2PCF covariances (Eqs.\,\ref{eq: cov2x2 Legendre terms}) in the limit $X=Y=Z=W$, and may be computed in the same manner as the $(r,\mu)$-space multi-field 2PCF covariance integrals, described in \citet[Sec.\,6]{rascalC}. Notably the inclusion of multiple fields gives two distinct Gaussian 4-point terms, which involve $\xi^{XZ}_{ik}\xi^{YW}_{jl}$ and $\xi^{XW}_{il}\xi^{YZ}_{ik}$, unlike the single $2\xi_{ik}\xi_{jl}$ term present previously. As in previous work, for a two-field scenario (denoted $S$ and $T$), if we compute the $C^{ST,TS}$ term (and its symmetries) as the average of $C^{ST,TS}$ and $C^{ST,ST}$, we can assume this symmetry to hold, greatly expediting computation.

\subsection{Comparison with Mock Legendre Covariances}\label{subsec: 2pcf-data-comparison}
The integrals of Sec.\,\ref{subsec: 2pcf_covariance} and \ref{subsec: 2pcf-covariance-multi} may be computed with small modification to the algorithm described in \citet{rascalC}, and have been implemented into the \texttt{RascalC} pipeline for both one and two sets of tracer particles. To assess whether our algorithm and covariance matrix formalism is producing correct results, we use the \texttt{RascalC} code to compute a 2PCF covariance matrix in Legendre polynomial bins, which may be compared to a sample covariance derived from a set of Quick Particle Mesh \citep[QPM;][]{2014MNRAS.437.2594W} mocks created for the CMASS-N survey from Data Release $12$ of the Baryon Oscillation Spectroscopic Survey \citep[BOSS;][]{2015ApJS..219...12A,2017MNRAS.470.2617A}, part of the Sloan Digital Sky Survey III \citep[SDSS-III;][]{2011AJ....142...72E}. In this test, the input 2PCF (used in the covariance integrals) is computed from the mean of 1000 finely-binned QPM \citep{2014MNRAS.437.2594W} mock galaxy correlation functions. All even multipoles up to $\ell = 6$ are used, with a covariance matrix radial binning scheme of $r\,\in[40,180]h^{-1}\,\mathrm{Mpc}$ and $\Delta r = 4h^{-1}\mathrm{Mpc}$, giving a total of 35 radial bins and four Legendre bins. Here, we use a random particle file provided by BOSS for clustering analyses, which uses $N_\mathrm{rand}=10N_\mathrm{gal}=6420510$ particles, and we adopt FKP \citep{1994ApJ...426...23F} weights. Before running the main code, $RR$ pair counts are computed using \texttt{corrfunc} \citep{2017ascl.soft03003S} for 35 radial bins (as above) and 100 angular bins which are fit to the boundless infinite survey model to determine the correction function $\Phi$ for each radial bin as a function of $\mu$ (cf.\,Eq.\,\ref{eq: RR_approximation}), using $V\overline{(nw)^2}=7.07h^3\,\mathrm{Mpc}^{-3}$. In radial bin $a$, $\Phi$ is parametrized as a polynomial function;
\beq
    \Phi^a(\mu) = \begin{cases} \beta_0^a + \beta_1^a\mu + \beta_2^a\mu^2 & \text{ if }0\geq\mu\geq0.75\\
                                \gamma_0^a + \gamma_1^a\mu + \gamma_2^a\mu^2 + \gamma_3^a\mu^3 & \text{ else,}\end{cases}
\eeq
with three components set by ensuring continuity and smoothness up to $d^2\Phi^a/d\mu^2$ at $\mu=0.75$. The piecewise function is adopted to account for the different behaviour of $\mu$ near unity, due to RSD effects. The free components $\{\beta^a\},\{\gamma^a\}$ are here fit using \texttt{scipy}. Following this, the covariance integrals are estimated using a total of $8\times 10^{12}$ quads of particles, taking $\sim 200$ CPU-hours to compute. As in previous works, non-Gaussianity is encapsulated via the inclusion of the shot-noise rescaling parameter, $\alpha$, calibrated using jackknife matrices. As previously mentioned, the above conversions of $(r,\mu)$ binning to Legendre binning do not apply for jackknife matrices, thus we adopt the value $\alpha=1.032$ for this mock, as found in \citet{rascalC}.

For the sample covariance matrix, we estimate the Legendre-moments of the anisotropic 2PCF, $\hat{\xi}_\ell(r)$, for 99 individual QPM mocks, using pair counts computed for $\Delta\mu = 1/120$ and $\Delta r = 4h^{-1}\,$Mpc. Given a radial bin $a$ and a set of $\mu$-bins spanning $[0,1]$ with centers at $\{\mu_c\}$, we estimate the Legendre-binned 2PCF in mock $n$ as
\beq\label{eq: qpm_legendre_approx}
    \hat{\xi}_\ell^{(n),a} \approx (2\ell+1)\Delta \mu \sum_c \hat\xi^{(n),a}_c L_\ell(\mu_c)\,,
\eeq
which becomes exact in the limit $\Delta\mu\rightarrow0$ (and is simply the discretized form of Eq.\,\ref{eq: Legendre-from-angular-2PCF-v2}). The sample covariance matrix in radial bins $a,b$ and Legendre moments $p,q$ is computed as
\beq\label{eq: qpm_covariance}
    \hat C^{ab}_{pq} = \operatorname{cov}(\hat\xi^a_p,\hat\xi^b_q) = \frac{1}{N_\mathrm{mocks}-1}\sum_{n=1}^{N_\mathrm{mocks}} \left[\left(\hat\xi^{(n),a}_p-\bar\xi^a_p\right)\left(\hat\xi^{(n),b}_q-\bar\xi^b_q\right)\right]
\eeq
where $\bar\xi$ indicates the mean over all $99$ mocks. Despite the large number of $\mu$-bins used here, we do not necessarily expect perfect agreement between our code and the QPM mocks here as the 2PCF greatly changes as $\mu\rightarrow1$, leading to non-negligible errors due to the finite binning. 

\begin{figure}
\centering
\begin{minipage}[t]{.45\textwidth}
  \includegraphics[width=\textwidth]{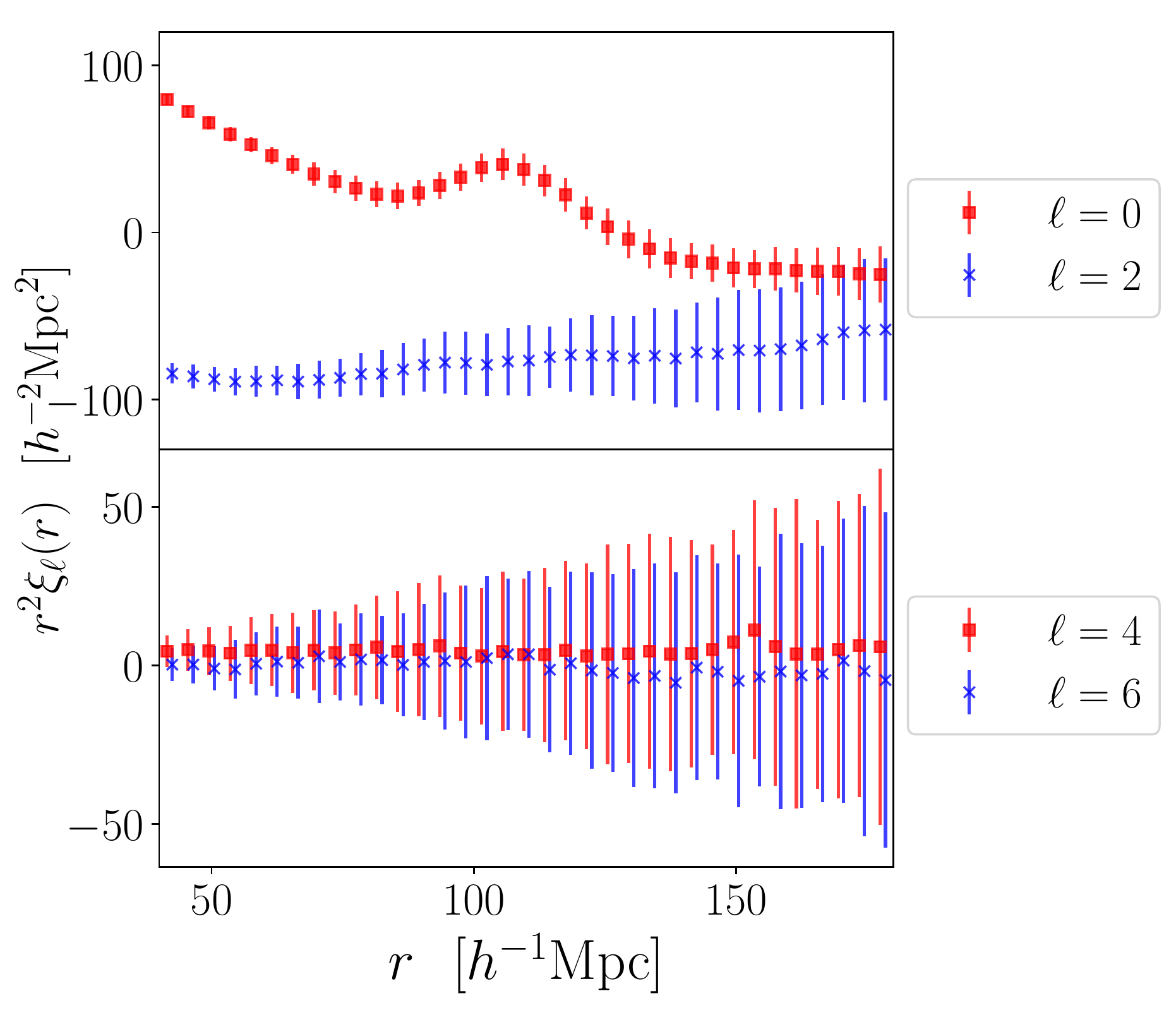}
\caption{Measured two-point correlation function (2PCF) multipoles from 99 Quick Particle Mesh \citep[QPM;][]{2014MNRAS.437.2594W} BOSS DR12 mocks, computed with the \texttt{corrfunc} code \citep{2017ascl.soft03003S}. In each mock, we measure the 2PCF from $\sim 6\times 10^5$ using the \citet{1993ApJ...412...64L} estimator in a total of 35 radial bins and 100 angular bins, before conversion into Legendre polynomial space via Eq.\,\ref{eq: Legendre-from-angular-2PCF-v2}. The displayed errors (computed as the standard deviation of the 99 2PCF measurements) represent the precision achievable from a single galaxy survey, and a small lateral displacement is added to the $\ell = 2$ and $6$ data-sets for visibility. We note highly significant monopole and quadrupole detections but little power in the higher multipoles.}
    \label{fig: 2PCF-Legendre-estimates}
\end{minipage}%
\hfill
\begin{minipage}[t]{.53\textwidth}
  \centering
  \includegraphics[width=0.8\textwidth]{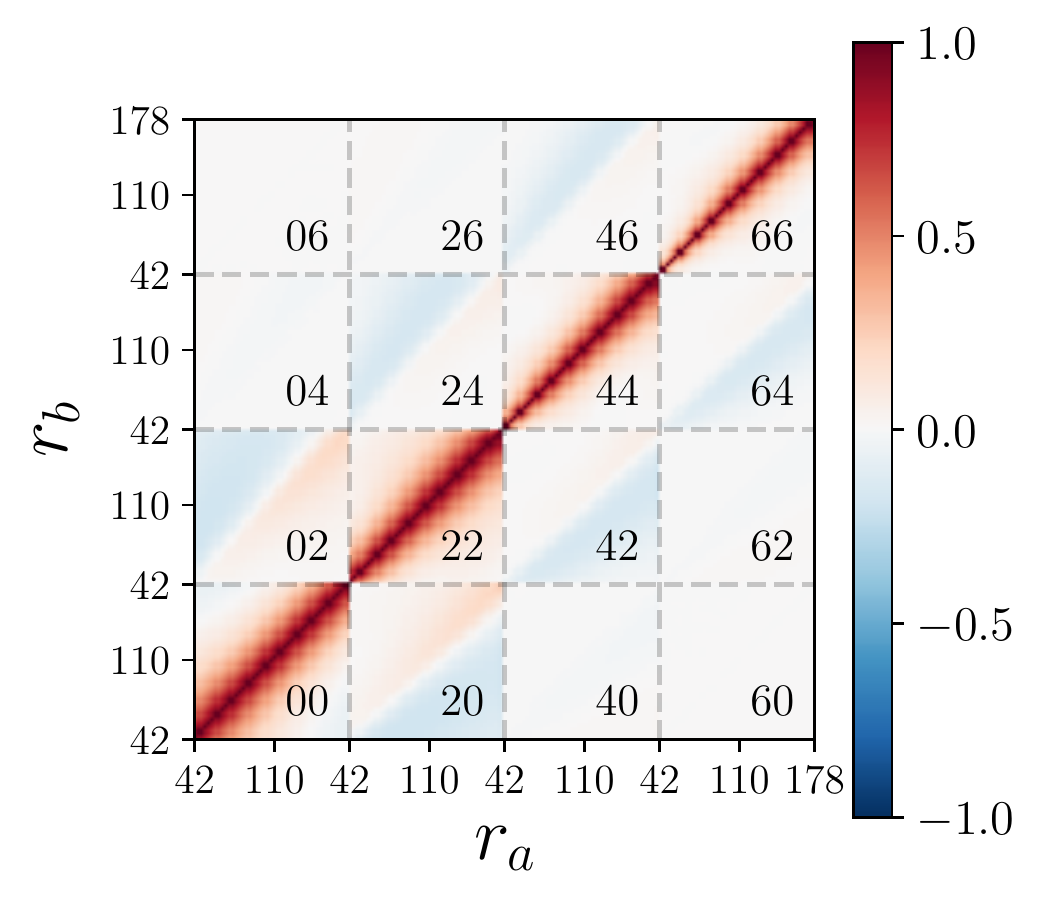}
\caption{Theoretical correlation matrix (Eq.\,\ref{eq: correlation_matrix}) for the anisotropic 2PCF estimates displayed in Fig.\,\ref{fig: 2PCF-Legendre-estimates}. Individual pixels represent the correlations at a pair of radial bins $\{a,b\}$, and we display results for 16 pairs of Legendre multipoles, $\{p,q\}$, with each pair in a separate sub-matrix, as indicated by the text. This matrix is computed with the \texttt{RascalC} code in $\sim 200$ CPU-hours, sampling covariance matrix integrals using a random particle catalog appropriate for the BOSS DR12 CMASS-N geometry \citep{2017MNRAS.470.2617A}. The estimator uses the mean correlation function estimated from 1000 QPM mock galaxy surveys, and non-Gaussianity is incorporated by way of a shot-noise rescaling parameter $\alpha=1.032$, as found by \citet{rascalC}. As described in the main text, the theoretical matrix has a noise level from numerical errors equivalent to that of $\sim 6\times 10^6$ mocks, though the error from systematics may be significantly larger. We note strongest correlations for identical multipoles which are weaker as $|p-q|$ increases.}
    \label{fig: smooth_correlation_matrix}
\end{minipage}%
\end{figure}

Before comparing theory and mock observations, we first we consider the structure of the covariance matrix estimate. In Fig.\,\ref{fig: smooth_correlation_matrix}, the correlation matrix $R^{ab}_{pq}$ is displayed for all even Legendre moment pairs $\{p,q\}$ up to the sixth moment, defining 
\beq\label{eq: correlation_matrix}
R^{ab}_{pq} = C^{ab}_{pq}/\sqrt{C^{aa}_{pp}C^{bb}_{qq}}.
\eeq
Notably, we observe strong positive correlations between bins with $p=q$, which are largest for $r_a\approx r_b$. These grow weaker both as $|r_a-r_b|$ and $|p-q|$ increase, with the $p=q+2$ submatrices displaying weakly positive correlations for $r_a<r_b$ and negative else. Analysis of the $C_{ab}^{pq}$ matrix itself shows small covariances for $\{p,q\} = \{0,0\}$, which grow greater for higher $\ell$, reflecting the increased difficulty in observing higher multipoles.

For comparison of theory and simulation, we require precision matrices, which, in the limit of zero noise (or $N_\mathrm{mocks}\rightarrow\infty$), are defined simply as covariance matrix inverses. As shown in \citet{wishart1928} and \citet{2007A&A...464..399H}, for finite $N_\mathrm{mocks}$, we must instead use the (compressed) sample precision matrix definition
\beq\label{eq: wishart-precision}
    \mathbf{\Psi}_\mathrm{sample} &=& (1-D)\hat{\mathbf{C}}_\mathrm{sample}^{-1}\\\nonumber
    D &=& \frac{N_\mathrm{bins}+1}{N_\mathrm{mocks}-1}
\eeq
to avoid a non-negligible bias. For the theoretical covariance matrix, which is not expected to obey Wishart statistics, we take a different approach, made possible by obtaining a set $\{\mathbf{C}_i\}$ of $N_\mathrm{samples}$ independent theoretical covariance matrix estimates. The bias-corrected precision is then defined as;
\beq\label{eq: non-wishart-precision}
    \mathbf{\Psi}_\mathrm{theory} &=& (\mathbb{I}-\tilde{\mathbf{D}})\hat{\mathbf{C}}^{-1}\\\nonumber
    \tilde{\mathbf{D}} &=& \frac{N_\mathrm{samples}-1}{N_\mathrm{samples}}\left[-\mathbb{I}+\frac{1}{N_\mathrm{samples}}\sum_i\hat{\mathbf{C}}_{[i]}^{-1}\hat{\mathbf{C}}_i\right] 
\eeq
\citep{2019MNRAS.487.2701O}, where $\hat{\mathbf{C}}_{[i]}$ is the mean of the covariance matrix estimates, excluding $\hat{\mathbf{C}}_i$. Since the \texttt{RascalC} algorithm is fully stochastic, we can easily compute independent matrix estimates by running the algorithm multiple times. The $\tilde{\mathbf{D}}$ matrix may also be used to assess the matrix convergence via the `effective number of mocks', defined by
\beq\label{eq: N_eff_computation}
    N_\mathrm{eff} &=& \frac{N_\mathrm{bins}+1}{|\tilde{\mathbf{D}}|^{1/N_\mathrm{bins}}}+1\nonumber
\eeq
\citep[cf.][Sec.\,3.4]{rascalC}. Here, using $N_\mathrm{samples}=20$, we obtain $N_\mathrm{eff} = 6\times 10^6$, implying that our matrix has the same noise level as one constructed from six million mocks, greatly subdominant to the Wishart noise in the QPM covariance created from 99 mocks. We caution that $N_\mathrm{eff}$ purely parametrizes the noise level from the numerical sampling. There is an additional (non-stochastic) source of error from the approximations made in our theoretical model (i.e. that the higher-point correlation functions can be modelled by increased shot-noise), which is more difficult to estimate. Here we probe this by comparing our estimates to those from mocks, though it is not certain how well the non-Gaussianity of the mocks themselves represents our Universe.

To assess the systematic difference between the Legendre matrix estimates, we use the `discriminant' matrix
\beq\label{eq: comparison_matrix}
    \mathbf{P} = \sqrt{\mathbf{\Psi}_\mathrm{RascalC}}^T \mathbf{C}_\mathrm{QPM} \sqrt{\mathbf{\Psi}_\mathrm{RascalC}} - \mathbb{I}
\eeq
(suppressing matrix components for clarity), where $\mathbf{\Psi}_\mathrm{RascalC}$ is the theoretical precision matrix (from Eq.\,\ref{eq: non-wishart-precision}), $\mathbf{C}_\mathrm{QPM}$ is the QPM covariance, $\mathbb{I}$ is the identity matrix and the square-root indicates the lower Cholesky decomposition. Note that we invert only the theoretical matrix here, since $N_\mathrm{mocks}$ is low, leading to a severely biased sample precision matrix. The discriminant matrix is plotted in Fig.\,\ref{fig: diff_matrix}, and any systematic deviations from unity indicate differences between the approaches. In this instance, we do not observe any obvious deviations between the matrices given the level of noise. Indeed, the matrix is found to have a mean and root-mean-square value of $0.05\%$ and $13\%$ respectively, which matches the expected deviations of $N_\mathrm{DoF}/\sqrt{N_\mathrm{mocks}}\sim 14\%$ from noise alone, for $N_\mathrm{DoF} = 140$ degrees of freedom. At higher $\ell$, there is are slightly increased deviations along the main diagonal, which we attribute to the increasing error arising the $\mu$-binning at larger Legendre moment. The clear strong agreement between the matrices indicates that our $RR$ window-correction function approximation has been successful and that our algorithm works as expected.

\section{Estimating the Three Point Correlation Function}\label{sec: 3PCF}
\subsection{3PCF Estimator in Angular Bins}
The above formalism maybe generalized to the isotropic three-point correlation function (3PCF), $\zeta$. The standard Szapudi-Szalay estimator gives
\beq\label{eq: Szapudi-Szalay}
    \zeta(r_1,r_2,\chi) = \frac{NNN(r_1,r_2,\chi)}{RRR(r_1,r_2,\chi)}
\eeq
\citep{1998ApJ...494L..41S} for $N = D - R$ as before, parametrizing the 3PCF via the lengths of two sides ($r_1$ and $r_2$) and the cosine of the angle between them, $\chi$.\footnote{In this paper $\mu$ refers to cosines of angles between a pair of particles and the LoS and $\chi$ refers to triangle internal angles. Unlike for $\mu$, the sign of $\chi$ is important here, thus we do not take the modulus.} Given radial bins $a, b$ and $\chi$-bin $c$, the binned triple counts are defined in small cells (each of which containing at most one particle) as
\beq\label{eq: NNN,RRR counts}
    NNN^{ab}_c &=& \sum_{i\neq j\neq k}n_in_jn_kw_iw_jw_k\delta_i\delta_j\delta_k\tilde\Theta_{ijk}^{ab,c}\\\nonumber
    RRR^{ab}_c &=& \sum_{i\neq j\neq k}n_in_jn_kw_iw_jw_k\tilde\Theta_{ijk}^{ab,c}.
\eeq
analogously to the $RR$ and $NN$ counts (Eq.\,\ref{eq: NN-RR-discrete}). The binning function $\tilde\Theta_{ijk}^{ab,c}$ selects triangles of the correct orientation via
\beq\label{eq: 3PCF-mu-kernel}
\tilde\Theta^{ab,c}_{ijk} = \begin{cases} 1 & \text{ if $|\vec r_{YX}|$ is in bin $a$ \textit{and} $|\vec r_{ZX}|$ is in bin $b$ \textit{and} $\chi_{YZ} = \frac{(\vec r_{YX})\cdot(\vec r_{ZX})}{|\vec r_{YX}||\vec r_{ZX}|}$ is in bin $c$}\\
        0 & \text{ else.}
    \end{cases}
\eeq
denoting $\vec r_{XY} = \vec r_X - \vec r_Y$ etc., where $X,Y,Z$ are some permutation of $\{i,j,k\}$. Importantly, this is symmetric under any permuation of the indices $i,j,k$, as are the summands in Eqs.\,\ref{eq: NNN,RRR counts}. We may equivalently write this as a product of univariate binning functions,
\beq\label{eq: 3point_binning}
   \tilde\Theta^{ab,c}_{ijk} = \left[\Theta^a(r_{ij})\Theta^b(r_{ik})\Theta^c(\chi_{jk}) + \text{ 5 perms.}\right],
\eeq
summing over the six distinct triangle configurations, with $r = |\vec r|$ and $\chi_{jk}$ being the cosine of the angle between the vectors $\vec r_j-\vec r_i$ and $\vec r_k-\vec r_i$. The full 3PCF estimator thus becomes
\beq\label{eq: 3PCF estimators}
    \hat\zeta^{ab}_c &=& \frac{1}{RRR_c^{ab}}\sum_{i\neq j\neq k}n_in_jn_kw_iw_jw_k\delta_i\delta_j\delta_k\tilde\Theta_{ijk}^{ab,c}.
\eeq

\subsection{Legendre Moment 3PCF Estimator}
The Legendre moments of the 3PCF may be computed in an analogous fashion to the 2PCF (following \citealt{2004ApJ...605L..89S,2015MNRAS.454.4142S}), by way of a survey-correction function (as in Sec.\,\ref{subsec: 2PCF-Legendre}). We first consider an alternative form for the $RRR^{ab}_c$ triple count in radial bins $(a,b)$ and small angular bin $\delta\chi$ (which is later shrunk to infinitesimal width). Due to the relabelling symmetry of Eq.\,\ref{eq: NNN,RRR counts} under permutations of $\{i,j,k\}$, all six $RRR$ terms arising from the different triangle configurations (Eq.\,\ref{eq: 3point_binning}) are identical, which gives 
\beq\label{eq: RRR_asym}
    RRR_c^{ab} = 6\sum_{i\neq j\neq k}n_in_jn_kw_iw_jw_k\Theta^a(r_{ij})\Theta^b(r_{ik})\Theta^c(\chi_{jk})
\eeq
(the $NNN$ estimator exhibits similar behavior). Converting the summation into an integral over the survey gives
\beq
RRR_c^{ab} = 6 \int d^3\vec x\,d^3 \vec y\,d^3\vec z\,n(\vec x)n(\vec y)n(\vec z)w(\vec x)w(\vec y)w(\vec z)\Theta^a(|\vec y-\vec x|)\Theta^b(|\vec z - \vec x|)\Theta^c(\chi_{(\vec z - \vec y)})
\eeq
where $\chi_{(\vec z - \vec y)}$ indicates the angle between the vectors $\vec z - \vec x$ and $\vec y - \vec x$. For an infinite uniform survey we can extract the factors of $n$ and $w$ and transform variables to $\vec t = \vec y - \vec x$ and $\vec u = \vec z - \vec x$;
\beq
    RRR_c^{ab}  = 6(nw)^3\left[\int d^3\vec x \right]\left[\int d^3\vec t\,d^3\vec u\,\Theta^a(|\vec t|)\Theta^b(|\vec u|)\Theta^c(\chi_{(\vec t - \vec u)})\right].
\eeq
The integral over $\vec x$ gives a factor of the survey volume $V$ and we may additionally rotate the $\vec u$ to $\tilde{\vec{u}}$ coordinate such that $\chi_{\tilde{\vec u}} = 0$ without affecting the radial parts, giving
\beq
    RRR_c^{ab} &\approx& 6V(nw)^3\int d^3\vec t\,d^3\tilde{\vec u}\,\Theta^a(|\vec t|)\Theta^b(|\tilde{\vec u}|)\Theta^c(\chi_{\vec t})\\\nonumber
    &=& 6V(nw)^3 \times 4\pi \int d\tilde u\, \Theta^a(\tilde u) \times 2\pi\int t^2dt\,\Theta^b(t)\int d\chi_t\,\Theta^c(\chi_t)\\\nonumber
    &\approx& 3V(nw)^3 \times v_a \times v_b\delta\chi
\eeq
In the final line, we have performed the integrals over the binning function (without restricting to positive $\chi$, unlike for the Legendre-binned 2PCF), recalling $v_a = 4\pi\left(r^3_{a,\mathrm{max}}-r^3_{a,\mathrm{min}}\right)/3$ (or $4\pi r_a^2\Delta r$ in the thin-bin limit). As for the $RR$ pair counts, we introduce a survey correction factor $\Phi(r_a,r_b,\chi)$ to make this expression exact for non-uniform and aperiodic surveys (itself now depending on two radial bins and an angle);
\beq\label{eq: triple_continuous}
RRR^{ab}(\chi)  \equiv \frac{3V\overline{(nw)^3}v_av_b}{\Phi(r_a,r_b,\chi)}
\eeq
also converting to a continuous function in $\chi$, with $\overline{(nw)^3}$ being the survey-averaged value of $(nw)^3$. Notably $\Phi(r_a,r_b,\chi)$ is symmetric under $a\leftrightarrow b$ interchange.

The Legendre moments of the 3PCF are defined as
\beq
    \hat\zeta_\ell^{ab} = \frac{2\ell+1}{2}\int_{-1}^1 d\chi\, \frac{NNN^{ab}(\chi)}{RRR^{ab}(\chi)}L_\ell(\chi)
\eeq
(cf.\,Eq.\,\ref{eq: Legendre-from-angular-2PCF}) where $L_\ell(\chi)$ is the Legendre polynomial of order $\ell$. Note that we must include both odd and even $\ell$ here (unlike for the 2PCF) due to the lack of $\chi\rightarrow-\chi$ symmetry. As before, we may write 
\beq
    NNN^{ab}(\chi) = \sum_c NNN^{ab}_c\Theta^c(\chi)/\delta\chi
\eeq
for inifinitesimal bin-size $\delta\chi$, which, combined with the continuous random triple count (Eq.\,\ref{eq: triple_continuous}), gives
\beq
    \hat\zeta_\ell^{ab} = \frac{2\ell+1}{6V\overline{(nw)^3}v_av_b}\frac{1}{\delta\chi}\int_0^1\,d\chi\,\sum_c \Theta^c(\chi)NNN_c^{ab}L_\ell(\chi)\Phi(r_a,r_b,\chi).
\eeq
Including the full estimator for $NNN$ (Eq.\,\ref{eq: NNN,RRR counts}) using an asymmetric binning function (cf.\,Eq.\,\ref{eq: RRR_asym}) gives
\beq
    \zeta_\ell^{ab} =  \frac{2\ell+1}{6V\overline{(nw)^3}v_av_b} \frac{1}{\delta\chi}\int_0^1\,d\chi\,\sum_c\Theta^c(\chi)L_\ell(\chi)\Phi(r_a,r_b,\chi)\times 6\sum_{i\neq j\neq k}\Theta^c(\chi_{jk})\Theta^a(r_{ij})\Theta^b(r_{ik})n_in_jn_kw_iw_jw_k\delta_i\delta_j\delta_k.
\eeq
As before, we integrate over $\chi$ in the $\delta\chi\rightarrow0$ limit, noting that this requires $\chi=\chi_{jk}$. Returning to a form symmetric under $\{i,j,k\}$ permutations, we obtain
\beq\label{eq: 3PCF-legendre-full-estimator}
    \hat\xi_\ell^{ab} = \frac{2\ell+1}{6V\overline{(nw)^3}v_av_b}\sum_{i\neq j\neq k}n_in_jn_kw_iw_jw_k\delta_i\delta_j\delta_k\tilde K^{ab,\ell}_{ijk}
\eeq
for symmetrized kernel function (cf.\,Eq.\,\ref{eq: 3PCF-mu-kernel})
\beq\label{eq: 3PCF-symmetrized-kernel}
    K^{ab,\ell}_{ijk} = \left[\Theta^a(r_{ij})\Theta^b(r_{ik})L_\ell(\chi_{jk})\Phi(r_a,r_b,\chi_{jk})+\text{ 5 perms.}\right].
\eeq
This is simply equal to $2L_\ell(\chi_{ab})\Phi(r_a,r_b,\chi_{ab})$ for a triangle with one side in bin $a$, one side in bin $b$ and opening angle $\chi_{ab}$. In practice, given a triplet of particles, we can compute contributions to the 3-point integral for each Legendre multipole in every $(a,b)$ pair of radial bins that the triplet falls into (with different radial bins arising from different $\{i,j,k\}$ permutations). A na\"{i}ve triple count over all particles in a survey using this estimator would count each particle six times; we include this degeneracy to simplify later covariance computations. The form presented above may be naturally extended to the anisotropic 3PCF estimator using spherical harmonic functions rather than Legendre polynomials in the kernel function, as in \citet{2018MNRAS.478.1468S}.

\section{Three-Point Correlation Function Covariance Matrices}\label{sec: 3PCF-covs}
\subsection{Cross Covariance of the 3PCF and 2PCF}\label{subsec: 3PCFx2PCF}
Given the anisotropic 2PCF and 3PCF estimators (Eqs.\,\ref{eq: 2PCF in Legendre Moments}\,\&\,\ref{eq: 3PCF-legendre-full-estimator}) we may construct the cross-covariance between them. We initially work in the Legendre basis, which is more efficient since it requires far fewer angular bins and is faster to compute. For radial bins $a,b$ ($c$) and Legendre moment $p$ ($q$) for the 3PCF (2PCF), the cross-covariance is defined as
\beq\label{eq: cov3x2-initial-def}
    \operatorname{cov}\left(\hat\zeta^{ab}_p\,,\,\hat\xi^{c}_q\right)\equiv C^{ab,c}_{p,q} = \av{\hat\zeta^{ab}_p \hat\xi^{c}_q} - \av{\hat\zeta^{ab}_p}\av{\hat\xi^{c}_q}.
\eeq
Inserting the relevant 2PCF and 3PCF estimators, we obtain
\beq\label{eq: cov3x2 estimator}
    C^{ab,c}_{p,q} &=& \frac{(2p+1)(2q+1)}{6V^2\overline{(nw)^3}\,\overline{(nw)^2}v_av_bv_c}\sum_{i\neq j\neq k}\sum_{l\neq m}n_in_jn_kn_ln_mw_iw_jw_kw_lw_m \times  K_{ijk}^{ab,p}\Theta^c(r_{lm})L_q(\mu_{lm})\Phi(r_c,\mu_{lm})\\\nonumber
    &\times& \left[\av{\delta_i\delta_j\delta_k\delta_m\delta_n}-\av{\delta_i\delta_j\delta_k}\av{\delta_m\delta_n}\right]
\eeq
We may expand this into three-, four- and five-point terms using the result
\beq\label{eq: cov3x2 splitting}
    \sum_{i\neq j\neq k\neq l\neq m}X_{ijk}Y_{lm} &=& \sum_{i\neq j\neq k\neq l\neq m}X_{ijk}Y_{lm} +6\sum_{i\neq j\neq k\neq l}X_{ijk}Y_{li} + 6\sum_{i\neq j\neq k}X_{ijk}Y_{ij}
\eeq
where $X_{ijk}$ and $Y_{lm}$ are totally symmetric under argument permutations.\footnote{This is obtained by noting that there are six ways to contract both one and two indices drawn from sets of two and three equivalent indices.} To proceed we must expand the expectations of the overdensity fields $\delta$ in the three-, four- and five-point terms. Denoting an $n$-point connected correlation function $\xi^{(n)}_{x_1...x_n}$ by $(x_1...x_n)$, we may use Wick's theorem to show
\beq
    \av{\delta_i\delta_j\delta_k\delta_l\delta_m}-\av{\delta_i\delta_j\delta_k}\av{\delta_l\delta_m} &=& (ijklm) + \left[(ij)(klm)+(ik)(jlm)+(jk)(ilm)\right]\\\nonumber
    &+&\left[(il)(jkm)+(im)(jkl)+(jl)(ikm)+(jm)(ikl)+(kl)(ijm)+(km)(ijl)\right]\\\nonumber
    \frac{n_i}{\alpha}\av{\delta_i\delta_j\delta_k\delta_l\delta_i} &=& \av{(1+\delta_i)(\delta_j\delta_k\delta_l)} = (ijkl)+(jkl)+\left[(ij)(kl)+(ik)(jl)+(il)(jk)\right]\\\nonumber
    \frac{n_in_j}{\alpha^2}\av{\delta_i\delta_j\delta_k\delta_i\delta_j} &=& \av{(1+\delta_i)(1+\delta_j)\delta_k} = (ijk)+(ik)+(jk)
\eeq
This has made use of the shot-noise contraction approximation of Eq.\,\ref{eq: shot-noise}. Further simplifications can be made via the relabelling symmetries of the summands under index permutation, for example, in the five-point term, $\xi_{ij}\zeta_{klm}$ may be transformed into $\xi_{ik}\zeta_{jlm}$ by relabelling $j\leftrightarrow k$, which does not change the summed quantity. In the expressions above, all quantities in square parentheses are similarly equal when we sum over particles. Inserting these contractions into the covariance matrix estimator (Eq.\,\ref{eq: cov3x2 estimator}) and splitting into three-, four- and five-point terms via Eq.\,\ref{eq: cov3x2 splitting} gives
\beq
    C^{ab,c}_{p,q} = {}^5 C^{ab,c}_{p,q}+\alpha\times {}^4 C^{ab,c}_{p,q} + \alpha^2\times {}^3 C^{ab,c}_{p,q}
\eeq
using the definitions (presented here as summations but easily convertible into integrals as before)
\beq
    {}^5 C^{ab,c}_{p,q} &=& \frac{(2p+1)(2q+1)}{6V^2\overline{(nw)^3}\,\overline{(nw)^2}v_av_bv_c}\sum_{i\neq j\neq k\neq l\neq m}n_in_jn_kn_ln_mw_iw_jw_kw_lw_m \times K^{ab,p}_{ijk}\Theta^c(r_{lm})L_q(\mu_{lm}) \Phi(r_a,\mu_{lm}) \\\nonumber
    &\times& \left[\xi^{(5)}_{ijklm}+3\,\xi_{ij}\zeta_{klm}+6\,\xi_{il}\zeta_{jkm}\right]\\\nonumber
    {}^4 C^{ab,c}_{p,q} &=& 6\times\frac{(2p+1)(2q+1)}{6V\overline{(nw)^3}\,\overline{(nw)^2}v_av_bv_c}\sum_{i\neq j\neq k\neq l}n_in_jn_kn_l\left(w_i\right)^2w_jw_kw_l \times K^{ab,p}_{ijk}\Theta^c(r_{il})L_q(\mu_{li})\Phi(r_c,\mu_{li})\\\nonumber
    &\times& \left[\xi^{(4)}_{ijkl}+\zeta_{jkl}+3\,\textcolor{blue}{\xi_{ij}\xi_{kl}}\right]\\\nonumber
    {}^3 C^{ab,c}_{p,q} &=& 6\times \frac{(2p+1)(2q+1)}{6V^2\overline{(nw)^3}\,\overline{(nw)^2}v_av_bv_c}\sum_{i\neq j\neq k}n_in_jn_k\left(w_iw_j\right)^2w_k\times K_{ijk}^{ab,p}\Theta^c(r_{ij})L_q(\mu_{ij})\Phi(r_c,\mu_{ij})\\\nonumber
    &\times& \left[\zeta_{ijk}+2\,\textcolor{blue}{\xi_{ik}}\right] 
\eeq
with the Gaussian terms marked in blue. Notably, there is no Gaussian contribution to the five-point term here, thus all Gaussian components of this covariance arise from shot-noise contractions. If there exists some model (or function interpolated from data) for the 3PCF, the full covariance may be computed by drawing sets of five particles, and computing their contributions to the relevant bins. Each set of five particles will contribute to up to three pairs of 3PCF radial bins $a,b$ (due to the three triangle sides), a single 2PCF radial bin $c$ and all required Legendre moment pairs $p,q$. If we are interested in computing only the Gaussian contribution, we need only draw sets of four particles, significantly expediting computation.

If instead we desired this correlation functions directly in $\mu$-bins rather than Legendre multipoles (used in clustering wedges e.g. \citealt{2012MNRAS.419.3223K,2017MNRAS.464.1640S}), we may use a similar expansion as above, with a slightly modified summand. This gives the expression (for internal triangle angle $\chi$-bin $d$ and pairwise LoS $\mu$-bin $e$)
\beq
    \operatorname{cov}\left(\hat\zeta^{ab}_d\,,\,\hat{\xi}^{c}_e\right) &=& {}^5 C^{ab,c}_{d,e}+\alpha\times {}^4 C^{ab,c}_{d,e} + \alpha^2\times {}^3 C^{ab,c}_{d,e}\\\nonumber
    {}^5C^{ab,c}_{d,e} &=& \frac{1}{RRR^{ab}_dRR^c_e}\sum_{i\neq j\neq k\neq l\neq m}n_in_jn_kn_ln_mw_iw_jw_kw_lw_m\tilde\Theta_{ijk}^{ab,d}\Theta^c(r_{lm})\Theta^e(\mu_{lm})\\\nonumber
    &\times& \left[\xi^{(5)}_{ijklm}+3\xi_{ij}\zeta_{klm}+6\zeta_{ijl}\xi_{km}\right]\\\nonumber
    {}^4C^{ab,c}_{d,e} &=& \frac{6}{RRR^{ab}_dRR^c_e}\sum_{i\neq j\neq k\neq l}n_in_jn_kn_l\left(w_i\right)^2w_jw_kw_l\tilde\Theta_{ijk}^{ab,d}\Theta^c(r_{li})\Theta^e(\mu_{li})\\\nonumber
    &\times& \left[\xi^{(4)}_{ijkl}+\zeta_{jkl}+3\textcolor{blue}{\xi_{ij}\xi_{kl}}\right]\\\nonumber
    {}^3C^{ab,c}_{d,e} &=& \frac{6}{RRR^{ab}_dRR^c_e}\sum_{i\neq j\neq k}n_in_jn_k\left(w_iw_j\right)^2w_k\tilde\Theta_{ijk}^{ab,d}\Theta^c(r_{ij})\Theta^e(\mu_{ij})\\\nonumber
    &\times& \left[\zeta_{ijk}+2\,\textcolor{blue}{\xi_{ik}}\right]
\eeq
which may be computed in a similar fashion to the Legendre-binned function, except that one would add contributions to a single pair of angular bins for a given set of five particles and triangle configuration, rather than all combinations of multipoles.

\subsection{Auto-covariance of the 3PCF}
\subsubsection{Theoretical Estimators}\label{subsec: 3PCF-auto}
The auto-covariance of the 3PCF, $\zeta$, may be found analogously to the above, involving decomposition into three-, four-, five- and six-point summations (or integrals in the continuous limit). Using $a,b$ ($c,d$) to denote the radial bins and $p$ ($q$) the Legendre multipole of the first (second) isotropic 3PCF, we obtain the expression
\beq\label{eq: cov3x3}
    \operatorname{cov}\left(\hat\zeta^{ab}_p\,,\,\hat\zeta^{cd}_q\right)&\equiv& C^{ab,cd}_{p,q} = \av{\hat\zeta^{ab}_p \hat\zeta^{cd}_q} - \av{\hat\zeta^{ab}_p}\av{\hat\zeta^{cd}_q}\\\nonumber
     &=& \frac{(2p+1)(2q+1)}{\left[6V\overline{(nw)^3}\right]^2v_av_bv_cv_d}\sum_{i\neq j\neq k}\sum_{l\neq m\neq n}n_in_jn_kn_ln_mn_nw_iw_jw_kw_lw_mw_n \times  K_{ijk}^{ab,p}K_{lmn}^{cd,q}\\\nonumber
    &\times& \left[\av{\delta_i\delta_j\delta_k\delta_l\delta_m\delta_n}-\av{\delta_i\delta_j\delta_k}\av{\delta_l\delta_m\delta_n}\right]
\eeq
using the 3PCF definition and Legendre kernel function of Eqs.\,\ref{eq: 3PCF-legendre-full-estimator}\,\&\,\ref{eq: 3PCF-symmetrized-kernel}. Note that the summand is symmetric under any permutation of $\{i,j,k\}$ or $\{l,m,n\}$. As before, this may be expanded with the identity
\beq\label{eq: cov3x3 expansions}
    \sum_{i\neq j\neq k}\sum_{l\neq m\neq n}X_{ijk}Y_{lmn} = \sum_{i\neq j\neq k\neq l\neq m\neq n}X_{ijk}Y_{lmn} + 9\sum_{i\neq j\neq k\neq l\neq m}X_{ijk}Y_{klm} + 18\sum_{i\neq j\neq k\neq l}X_{ijk}Y_{jkl}+6\sum_{i\neq j\neq k}X_{ijk}Y_{ijk}
\eeq
for totally symmetric functions $X$ and $Y$, with the symmetry factors derived by considering the number of possible ways to contract one, two or three pairs of indices from two sets of three. This allows Eq.\,\ref{eq: cov3x3} to be expressed in terms of three-, four-, five- and six-point functions and the order of index contractions is chosen for later use. 

We must now consider the expansion of the six overdensity fields and their various contractions. For the sake of brevity, we combine terms with the same $\{i,j,k\}$ and $\{l,m,n\}$ permutation symmetries, since these all give identical contributions when we sum over all particles. The choice of index permutations shown have been chosen carefully for later convenience. From Wick's theorem
\beq
    \av{\delta_i\delta_j\delta_k\delta_l\delta_m\delta_n}_\mathrm{sym}-\av{\delta_i\delta_j\delta_k}\av{\delta_k\delta_m\delta_n} &=& (ijklmn) + 3(ij)(klmn) + 9(il)(jkmn)+ 3(mn)(ijkl)\\\nonumber
    &+& 3(mn)(ijkl) + 9(ijl)(kmn)+9(ij)(kl)(mn)+6(il)(jm)(kn)\\\nonumber
    \frac{n_k}{\alpha}\av{\delta_i\delta_j\delta_k\delta_l\delta_m\delta_k}_\mathrm{sym} = \av{(1+\delta_k)\delta_i\delta_j\delta_l\delta_m}_\mathrm{sym} &=& (ijklm)+(ijlm)+2(ik)(jlm)+2(kl)(ijm)+(ij)(klm)\\\nonumber
    &+&4(il)(jkm)+(lm)(ijk)+(ij)(lm)+2(il)(jm)\\\nonumber
    \frac{n_jn_k}{\alpha^2}\av{\delta_i\delta_j\delta_k\delta_l\delta_j\delta_k}_\mathrm{sym} = \av{(1+\delta_i)(1+\delta_j)\delta_k\delta_l}&=&(ijkl)+(il)(jk)+2(ij)(kl)+2(ijl)+(il)\\\nonumber
    \frac{n_in_jn_k}{\alpha^3}\av{\delta_i\delta_j\delta_k\delta_i\delta_j\delta_k}_\mathrm{sym} = \av{(1+\delta_i)(1+\delta_j)(1+\delta_k)}&=& (ijk)+3(ij)+1
\eeq
denoting $\xi^{(n)}_{x_1...x_n}$ by $(x_1...x_n)$ as before and using the shot-noise contraction identity (Eq.\,\ref{eq: shot-noise}). Inserting the above random field expansions into Eq.\,\ref{eq: cov3x3} and splitting into three- to six-point summations as in Eq.\,\ref{eq: cov3x3 expansions}, we obtain the full solution for the 3PCF auto-covariance matrix;
\beq\label{eq: cov3x3 full terms}
    C_{p,q}^{ab,cd} &=& {}^6C^{ab,cd}_{p,q} + \alpha\times{}^5C^{ab,cd}_{p,q} + \alpha^2\times{}^4C^{ab,cd}_{p,q} + \alpha^3\times{}^3C^{ab,cd}_{p,q}\\\nonumber
    {}^6C^{ab,cd}_{p,q} &=& \frac{(2p+1)(2q+1)}{\left[6V\overline{(nw)^3}\right]^2v_av_bv_cv_d}\sum_{i\neq j\neq k\neq l\neq m\neq n}n_in_jn_kn_ln_mn_nw_iw_jw_kw_lw_mw_nK^{ab,p}_{ijk}K^{cd,q}_{lmn}\\\nonumber
    &\times&\left[\xi_{ijklmn}^{(6)}+3\,\xi_{ij}\xi^{(4)}_{klmn}+9\,\xi_{il}\xi^{(4)}_{jkmn}+3\,\xi_{mn}\xi^{(4)}_{ijkl}+9\,\zeta_{ijl}\zeta_{kmn}+9\,\textcolor{blue}{\xi_{ij}\xi_{kl}\xi_{mn}}+6\,\textcolor{blue}{\xi_{il}\xi_{jm}\xi_{kn}}\right]\\\nonumber
    {}^5C^{ab,cd}_{p,q} &=& 9\times \frac{(2p+1)(2q+1)}{\left[6V\overline{(nw)^3}\right]^2v_av_bv_cv_d}\sum_{i\neq j\neq k\neq l\neq m}n_in_jn_kn_ln_mw_iw_j\left(w_k\right)^2w_lw_mK_{ijk}^{ab,p}K_{klm}^{cd,q}\\\nonumber
    &\times&\left[\xi^{(5)}_{ijklm}+\xi^{(4)}_{ijlm}+2\,\xi_{ik}\zeta_{jlm}+2\,\xi_{kl}\zeta_{ijm}+\xi_{ij}\zeta_{klm}+4\,\xi_{il}\zeta_{jkm}+\xi_{lm}\zeta_{ijk}+\textcolor{blue}{\xi_{ij}\xi_{lm}}+2\,\textcolor{blue}{\xi_{il}\xi_{jm}}\right]\\\nonumber
    {}^4C^{ab,cd}_{p,q} &=& 18\times \frac{(2p+1)(2q+1)}{\left[6V\overline{(nw)^3}\right]^2v_av_bv_cv_d}\sum_{i\neq j\neq k\neq l}n_in_jn_kn_lw_i\left(w_jw_k\right)^2w_lK_{ijk}^{ab,p}K_{jkl}^{cd,q}\\\nonumber
    &\times& \left[\xi^{(4)}_{ijkl}+2\,\zeta_{ikl}+\textcolor{blue}{\xi_{il}\xi_{jk}}+2\,\textcolor{blue}{\xi_{ij}\xi_{kl}}+\textcolor{blue}{\xi_{il}}\right]\\\nonumber
    {}^3C^{ab,cd}_{p,q} &=& 6\times \frac{(2p+1)(2q+1)}{\left[6V\overline{(nw)^3}\right]^2v_av_bv_cv_d}\sum_{i\neq j\neq k}n_in_jn_k\left(w_iw_jw_k\right)^2K_{ijk}^{ab,p}K_{ijk}^{cd,q}\\\nonumber
    &\times&\left[\zeta_{ijk}+3\,\textcolor{blue}{\xi_{ij}}+\textcolor{blue}{1}\right]
\eeq
where we color the Gaussian terms in blue as before, noting Gaussian contributions to all terms. This may also be computed in $\chi$-bins $(e,f)$ rather than Legendre moments $(p,q)$, which has the same functional form, except that the $K^{ab,p}K^{cd,q}$ kernels are replaced by $\tilde\Theta^{ab,e}\tilde\Theta^{cd,f}$ binning functions (as in Eq.\,\ref{eq: 3point_binning}) and the prefactor changed to $\left[RRR^{ab}_eRRR^{cd}_f\right]^{-1}$. 

\subsubsection{Covariance Computation}\label{subsec: 3pcf-algorithm}
Here we outline how the Gaussian contributions to the above 3PCF integral may be computed efficiently, via sampling sets of six spatial points (`hexes'), analogous to the quad-sampling algorithm introduced in \citet{rascalC} for the 2PCF auto-covariance. Although the integrals are of higher dimension than for the 2PCF, the use of Legendre polynomial bins and matrix compression (see Sec.\,\ref{subsec: BOSS-3PCF}) ensure that the number of bins is not too large, thus the matrices are relatively quick to converge. We begin by splitting up the Gaussian parts of Eqs.\,\ref{eq: cov3x3 full terms} into two distinct sets of summations giving

\beq\label{eq: cov3x3 splitting}
    {}^6C^{ab,cd}_{p,q} &=& 9\sum_{i\neq j\neq k\neq l\neq m\neq n}\Omega^{ab,p}_{ijk}\Omega^{cd,q}_{lmn}\left[\xi_{ij}\xi_{kl}\xi_{mn}\right]+6\sum_{i\neq j\neq k\neq l\neq m\neq n}\Omega^{ab,p}_{ijk}\Omega^{cd,q}_{lmn}\left[\xi_{il}\xi_{jm}\xi_{kn}\right]\\\nonumber
    {}^5C^{ab,cd}_{p,q} &=& 9\sum_{i\neq j\neq k\neq l\neq m}\Omega^{ab,p}_{ijk}\Omega^{cd,q}_{klm}\left[\xi_{ij}\xi_{lm}\right]+18\sum_{i\neq j\neq k\neq l\neq m}\Omega^{ab,p}_{ijk}\Omega^{cd,q}_{klm}\left[\xi_{il}\xi_{jm}\right]\\\nonumber
    {}^4C^{ab,cd}_{p,q} &=& 36\sum_{i\neq j\neq k\neq l}\Omega^{ab,p}_{ijk}\Omega^{cd,q}_{jkl}\left[2\xi_{ij}\xi_{kl}\right]+18\sum_{i\neq j\neq k\neq l}\Omega^{ab,p}_{ijk}\Omega^{cd,q}_{jkl}\left[\xi_{il}\xi_{jk}+\xi_{il}\right]\\\nonumber
    {}^3C^{ab,cd}_{p,q} &=& 3\sum_{i\neq j\neq k}\Omega^{ab,p}_{ijk}\Omega^{cd,q}_{ijk}\left[3\xi_{ij}+1\right]+3\sum_{i\neq j\neq k\neq l}\Omega^{ab,p}_{ijk}\Omega^{cd,q}_{ijk}\left[3\xi_{jk}+1\right],
\eeq
using the definition
\beq
\Omega_{ijk}^{ab,p} = \frac{2p+1}{6V\overline{(nw)^3}v_av_b}\times n_in_jn_kw_iw_jw_kK_{ijk}^{ab,p},
\eeq
and noting that we have rewritten the $18\xi_{ij}$ factor in the three-point term as $9\xi_{ij}+9\xi_{jk}$ (via interchange symmetry of the kernels and summations). These summations can simply be converted into integrals, e.g. replacing $\sum_i$ with $\int d^3\vec r_i$, and $\xi_{ij}$ with $\xi(\vec r_i-\vec r_j)$. The two separate integral terms shown in Eq.\,\ref{eq: cov3x3 splitting} are henceforth labelled $A$ and $B$ and are computed separately. As detailed in Appendix \ref{sec: 3PCF-cancellation}, the first of the six-point terms involves an integral over the full 2PCF $\xi(\vec r)$, constrained only by the survey geometry, which should be zero for a uniform ideal survey. In reality, we do not expect perfect cancellation, but we expect the term to be small if the size of the survey is large compared with the BAO scale and maximum radial bin. 

The two sets of terms may be computed by drawing sets of three to six points in three-dimensional space (subject to the survey selection functions) and computing the contribution pf each to the relevant summations. For efficiency, we draw these points from an input set of random particle positions (such as those provided for clustering random counts), which naturally encode the survey number density distribution and mask. Integral convergence is greatly improved via importance sampling, where we preferentially sample points in space which have large contributions to the integral, and divide by the probability that that set of points was chosen. As in \citet{rascalC}, we first assign particles to `sampling cells', each of which contain a few particles, allowing us to pre-compute the probability that a particular cell is chosen. To draw a random particle from a given primary, we simply pick a secondary cell and choose a random particle representative from within. 

Here, we consider two ways to draw a cell with center at $r_a$ from one at $r_b$; weighting by a $|\vec r_a-\vec r_b|^{-2}$ kernel or by the isotropic 2PCF $\xi(|\vec r_a-\vec r_b|)$. The former is appropriate when we wish to fill up radial bins evenly (e.g. for two legs in the same triangle) whilst the latter is used for integrands containing the anisotropic term $\xi_{ab}$ (although can also be used to fill radial bins since $\xi\sim r^{-2}$ at leading order). The sampling probabilities using these two kernels are those given in \citet[][Sec.\,3.4]{rascalC}. The sampling strategy was tested by comparing the 2PCF-independent 3-point term to semi-analytic predictions in the uniform periodic limit, and found to be in excellent agreement across the binning ranges used here. 

\begin{figure}
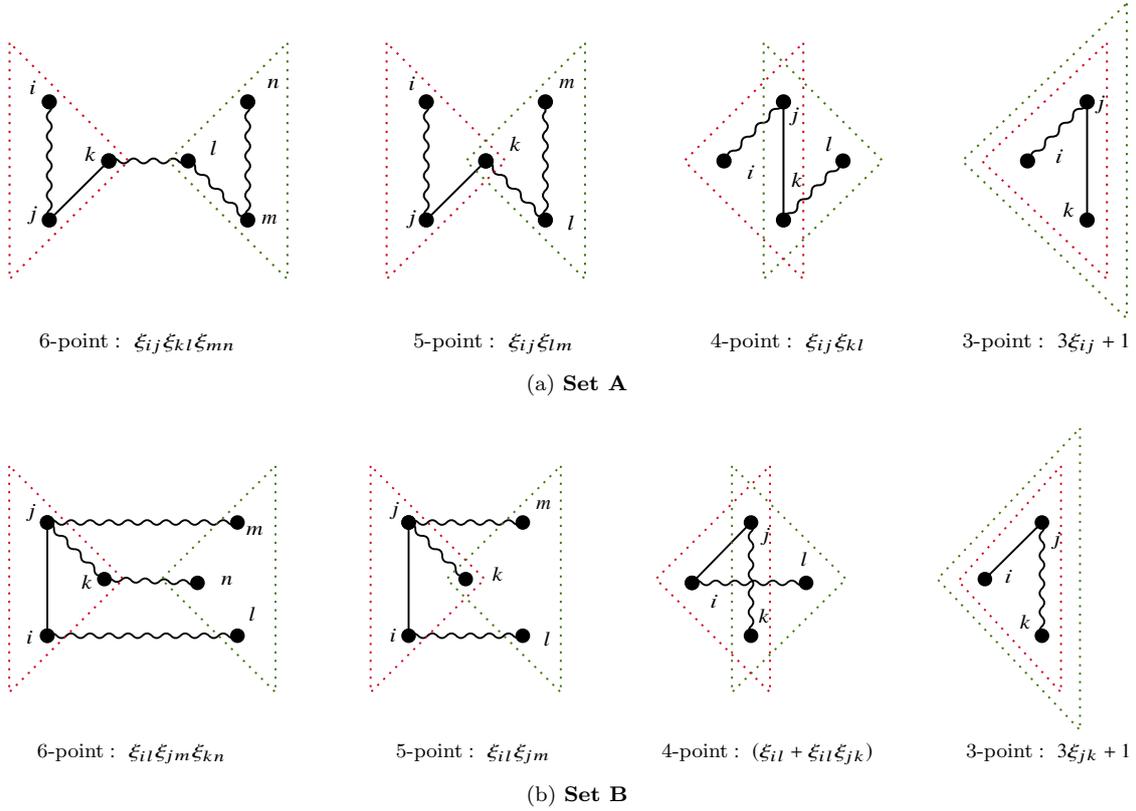
%
    \centering
    \subfloat[\textbf{Set A}]{{\includestandalone[width=.85\textwidth]{cov3x3-tikzA} }}%
    \\
    \subfloat[\textbf{Set B}]{{\includestandalone[width=.85\textwidth]{cov3x3-tikzB} }}%
    \caption{Pictorial representation of the sampling strategies used to compute the two sets of contributions to the 3PCF auto-covariance matrix integrals. The $d$-point covariance terms depend on $d$ points in space (for $d\in\{3,4,5,6\}$); we here depict how these are chosen, as well as which term each configuration corresponds to. To compute the integrals we first draw three points in space, $\{i,j,k\}$, then add the $l$, $m$ and $n$ points successively to find a single contribution to each integral. Here a straight line indicates that a point is drawn from the previous via a $1/r^2$ weighting (for separation $r$), with a wavy line indicating a draw via an isotropic $\xi(r)$ kernel. Each covariance integral depends on two triangles of points; the triplets of particles corresponding to each triangle are indicated by dashed lines.}%
    \label{fig: sampling_diagrams}%
\end{figure}

For both sets of integrals, we begin by drawing a single point from the set of random particles, then two more according to $r^{-2}$ and $\xi(r)$ kernels, and using these to compute the three-point integral contribution. Following this, a fourth particle is chosen from one of the previous particles via the $\xi(r)$ kernel and used to compute the four-point integral contribution, which is then repeated for the five- and six-point terms, before the process is repeated. Since the higher-point integrals are expected to take longer to converge, we usually sample multiple quads of points from each triple, multiple quints (sets of five points) from each quad and multiple hexes from each quint. For a given set of points (which form two triangles), we compute the contributions to all pairs of Legendre moments $p,q$ and all valid combinations of radial bins $a,b,c,d$ (using the triangle kernel of Eq.\,\ref{eq: 3PCF-symmetrized-kernel}). In Fig.\,\ref{fig: sampling_diagrams} we describe diagrammatically the sampling strategies used to compute the two sets of integrals, showing the particle from which each subsequent particle is sampled, along with the choice of kernel. The exact ordering is carefully chosen to ensure that the integrals converge as quickly as possible.

\subsubsection{Application to BOSS DR12 Mocks}\label{subsec: BOSS-3PCF}

\begin{figure}
\centering
\begin{minipage}[t]{.40\textwidth}
  \centering
  \includegraphics[width=.9\textwidth]{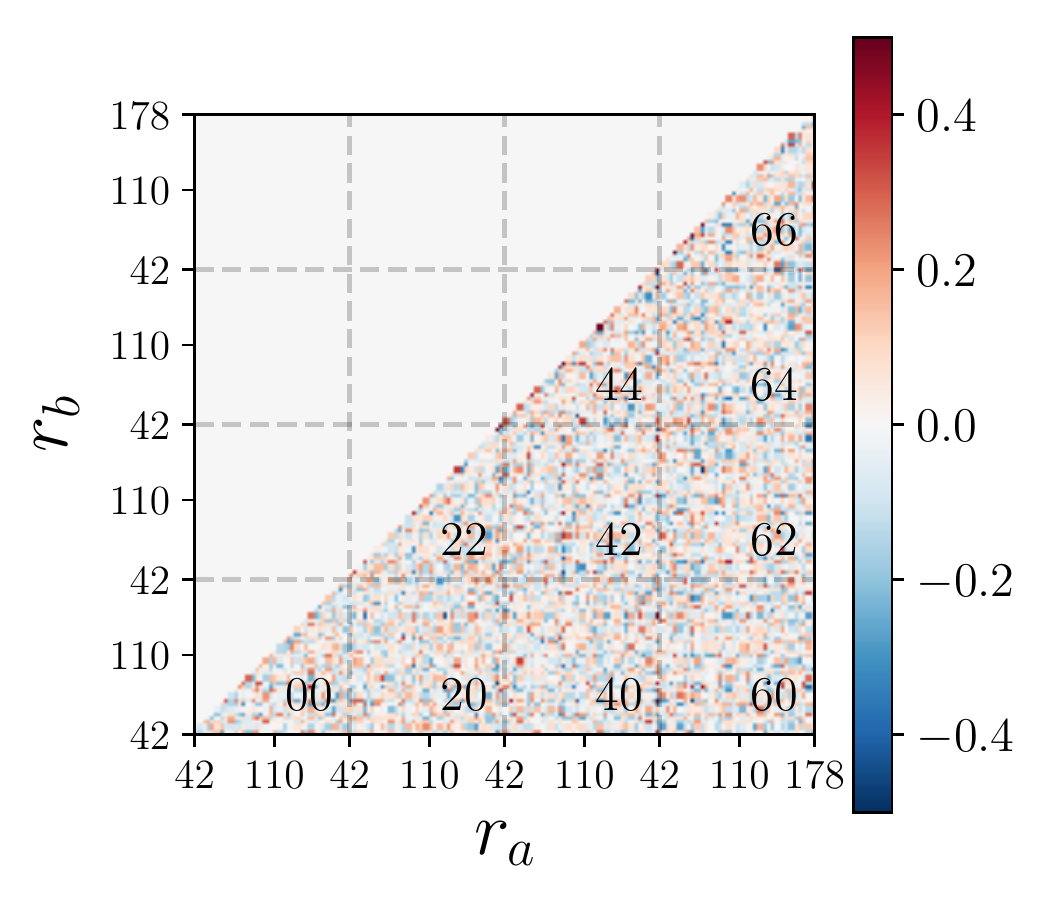}
  \caption{Discriminant matrix $\mathbf{P}$ between theoretical and sample covariance matrices of the anisotropic 2PCF in Legendre bins, defined in Eq.\,\ref{eq: comparison_matrix}. The theoretical matrix is displayed in reduced form in Fig.\,\ref{fig: smooth_correlation_matrix}, with the sample matrix being constructed from angularly-binned 2PCF estimates from 99 QPM mocks, as shown in Fig.\,\ref{fig: 2PCF-Legendre-estimates}. Pairs of multipoles are indicated by the text, as in Fig.\,\ref{fig: smooth_correlation_matrix}, and we note that $\mathbf{P}$ is zero for identical matrices in the absence of noise. Since the matrix is symmetric by definition, we only show the lower diagonal, to avoid spurious correlations being observed due to the symmetry. As described in Sec.\,\ref{subsec: 2pcf-data-comparison}, the difference between the matrices appears to be solely consistent with noise at this resolution level.}
  \label{fig: diff_matrix}
\end{minipage}
\hfill
\begin{minipage}[t]{.57\textwidth}
%\vspace{0pt}
\centering
  \includegraphics[width=.9\textwidth]{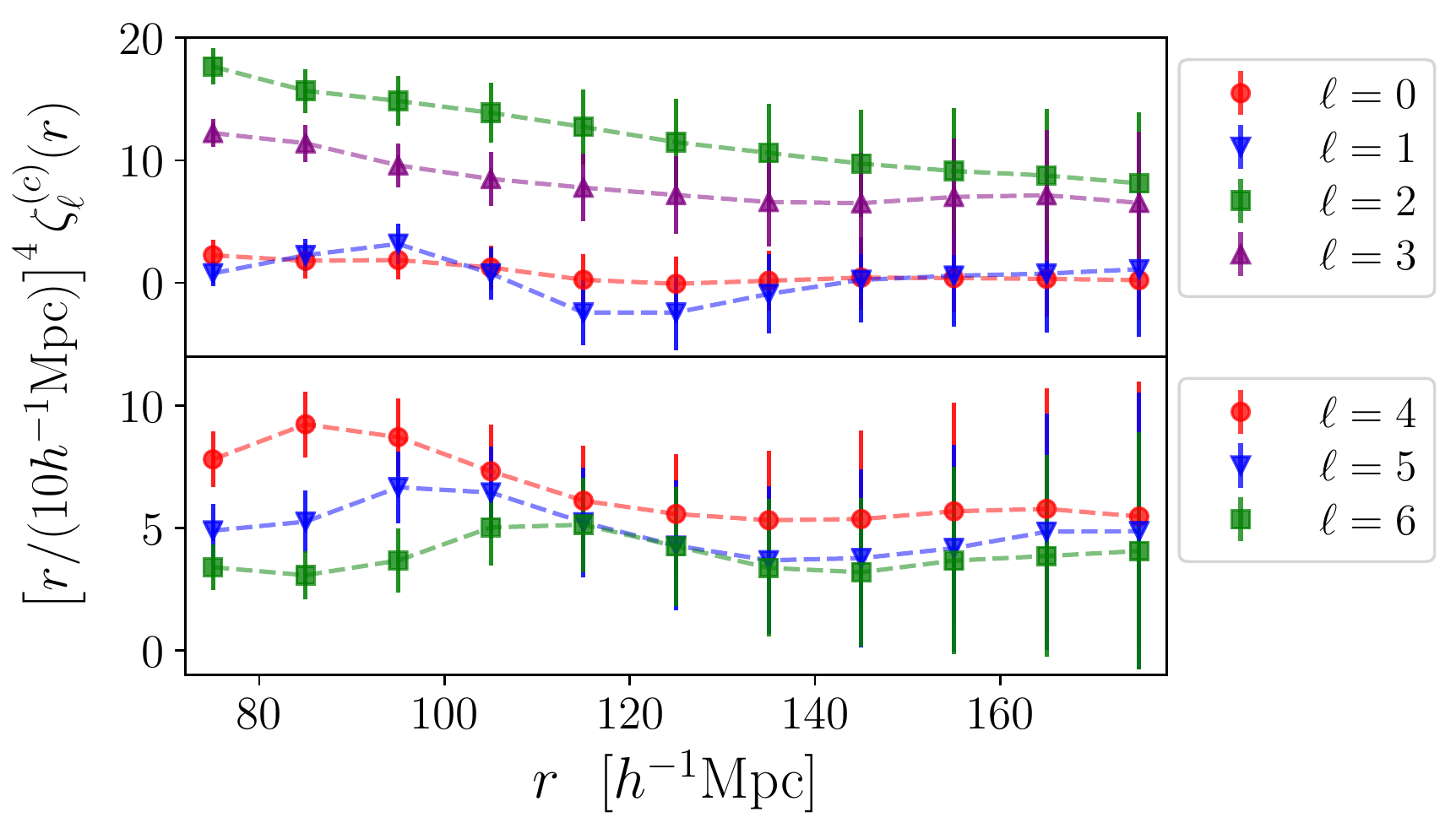}
\caption{Compressed three-point correlation function (3PCF) multipoles measured from 250 BOSS DR12 Quick Particle Mesh \citep[QPM;][]{2014MNRAS.437.2594W} mocks using the $\mathcal{O}(N^2)$ algorithm of \citet{2015MNRAS.454.4142S}. The 3PCF is measured from $\sim 6\times 10^5$ galaxies in each mock, using the \citet{1998ApJ...494L..41S} estimator to correct for boundary effects, and the compression of \citet{2015MNRAS.448....9S} to convert the 3PCF to a set of one-dimensional functions, averaging over radial bins which avoid the (non-Gaussian) squeezed limit. Error bars represent the standard deviation of the 250 measurements and are not normalized by $\sqrt{N_\mathrm{mocks}}$, i.e. they represent the precision possible from a single galaxy survey. There is a highly significant measurement of non-Gaussianity at smaller scales and we note broad consistency with the 3PCF measurements of \citet{2015arXiv151202231S}, computed for the \textsc{Multi-Dark Patchy} BOSS DR12 simulations. These measurements are used to compute the 3PCF covariances of Sec.\,\ref{subsec: BOSS-3PCF}.}
    \label{fig: 3PCF-measurements}
\end{minipage}
\end{figure}

To ensure that our theoretical 3PCF auto-covariance estimator is working correctly and to assess the validity of our shot-noise rescaling approximations, we must compare the output matrices to those from a suite of simulations. For this, we utilize QPM mocks appropriate for the BOSS DR12 surveys, as in Sec.\,\ref{subsec: 2pcf-data-comparison}. To form a sample covariance matrix, we first compute the 3PCF for each of 250 mocks using the algorithm of \citet{2015MNRAS.454.4142S}, which has complexity $\mathcal{O}(N^2)$ for $N$ galaxies, and incorporates the survey geometry via the multipoles of the $RRR$ triple count. We adopt a binning strategy of $\Delta r = 10h^{-1}\mathrm{Mpc}$ up to a maximum radius of $180h^{-1}\mathrm{Mpc}$ (matching \citealt{2015arXiv151202231S}) using all Legendre multipoles up to $\ell_\mathrm{max} = 6$. $\zeta_\ell$ multipoles are estimated as a ratio of $NNN$ and $RRR$ counts (cf.\,Eq.\,\ref{eq: Szapudi-Szalay}), requiring a large number of random galaxy positions to compute $N=D-R$ counts (which involve significant cancellation), yet far fewer for the $RRR$ counts, as noted in \citet{2015MNRAS.454.4142S}. We follow the former paper's approach of partitioning the random catalog into 32 random subsets, each with $N_\mathrm{rand}=1.5N_\mathrm{gal}$ particles, and averaging together the 32 resulting $NNN$ counts, giving $N_\mathrm{rand} = 48N_\mathrm{gal}$ in total.  The $RRR$ count is found to be well estimated using a single subset of randoms. For each QPM mock, 3PCF computation took $\sim 2$ hours on 32 parallel CPUs, requiring a total of $\sim 2\times 10^4$ CPU-hours for the complete analysis. 

For a reasonable choice of bin-width and $\ell_\mathrm{max}$, the total number of independent 3PCF bins (equal to $N_r(N_r+1)(\ell_\mathrm{max}+1)/2$ for $N_r$ radial bins) becomes large and hence difficult to interpret. To facilitate simpler visualization and covariance matrix comparison, we adopt the compression suggested by \citet{2015MNRAS.448....9S} and \citet{2015arXiv151202231S}, defining the compressed 3PCF, $\zeta_{(c)}$, as
\beq\label{eq: compressed-3PCF}
    \zeta^{a}_{(c),p} = \frac{\sum_{b\in S(a)}\zeta^{ab}_p v_b}{\sum_{b\in S(a)}v_b}\\\nonumber
\eeq
where $v_b$ is the volume defined by (radial) bin $b$ and $S(a)$ is the set of all bins $b$ satisfying $3\Delta r\leq r_{b,\mathrm{min}}< r_{a,\mathrm{min}}-3\Delta r$, chosen to avoid the (strongly non-Gaussian) squeezed limit of the 3PCF. For the binning strategy described above, compression restricts utilized bins to $r_b\geq30h^{-1}\mathrm{Mpc}$ and $r_a\geq70h^{-1}\mathrm{Mpc}$ (since there are no $b$ bins in $S(a)$ for smaller $r_a$), giving a total of $11$ bins in the compressed 3PCF for each Legendre multipole, compared with $15\times16\,/\,2=120$ in the uncompressed function. The resulting $\zeta^a_{(c),p}$ measurements for the 250 QPM mocks are displayed in Fig.\,\ref{fig: 3PCF-measurements} and we note highly significant measurements of the $\ell\geq2$ multipoles for moderate $r$, indicating clear non-Gaussianity. These results are broadly consistent with the equivalent measurements for the \textsc{MultiDark-Patchy} mocks \citep{2014MNRAS.439L..21K,2014arXiv1405.6273K,2015MNRAS.450.1836K} presented in \citet{2015arXiv151202231S} for the same survey geometry.

\begin{figure}
\centering
\begin{minipage}[t]{.53\textwidth}
  \centering
  \includegraphics[width=0.7\textwidth]{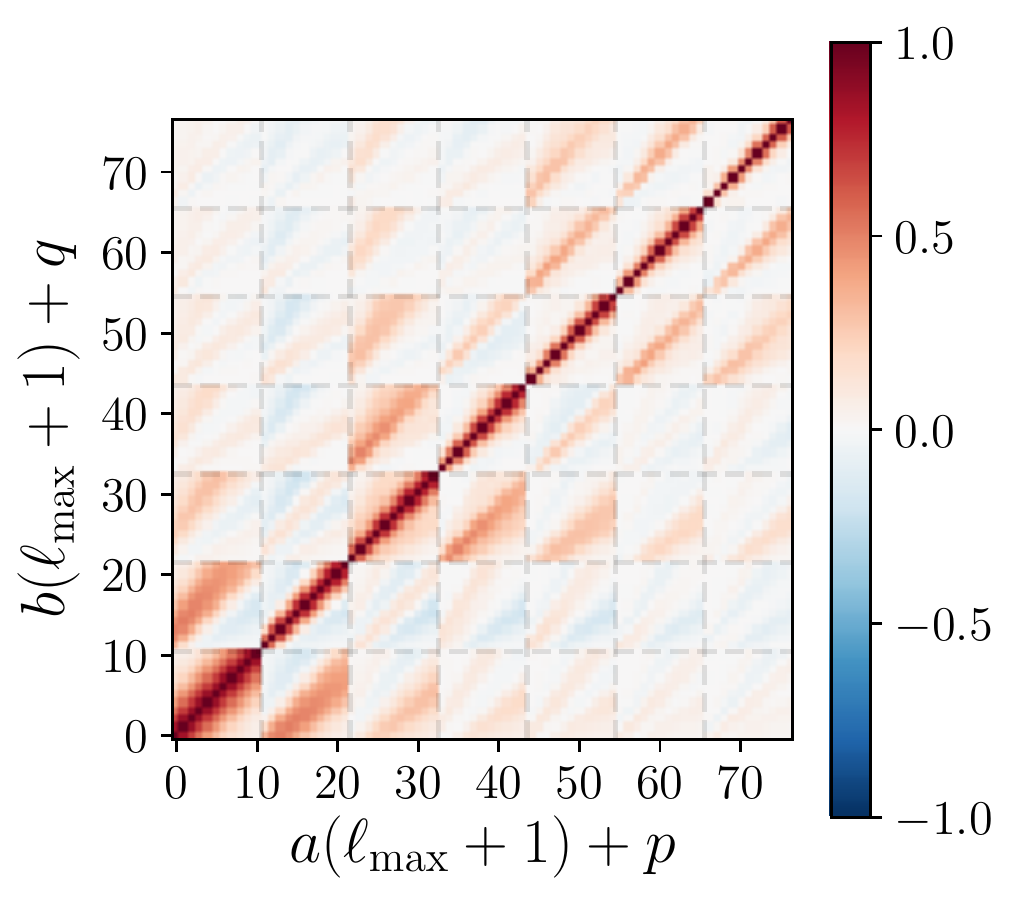}
    \caption{Theoretical correlation matrix (Eq.\,\ref{eq: correlation_matrix}) for the compressed isotropic 3PCF measurements presented in Fig.\,\ref{fig: 3PCF-measurements} for the BOSS DR12 CMASS-N geometry. This is computed using the newly-extended \texttt{RascalC} code in $\sim 200$ CPU-hours, with a numerical noise-level comparable to that of $2\times 10^4$ mocks. The compressed matrix is related to the full 3PCF covariance by Eq.\,\ref{eq: compressed-covariance}. This is analogous to Fig.\,\ref{fig: smooth_correlation_matrix}, except that we now use the 3PCF and display data for both odd and even Legendre multipoles, from $(p,q) = (0,0)$ (lower left) to $(6,6)$ (upper right). Each multipole pair inhabits its own submatrix, with the radial bin indices $a,b$ running from $0$ to $10$, denoting linear bins from $r=70-80h^{-1}\mathrm{Mpc}$ to $r=170-180h^{-1}\mathrm{Mpc}$. To encapsulate some degree of non-Gaussianity, we have included a shot-noise rescaling parameter of $\alpha = 1.20$, calibrated by minimizing the Kullback-Leibler \citep{kullback1951} divergence between the noisy sample matrix and the smooth theoretical matrix. We note strong positive correlations between bins with the same multipole, with significant negative correlations when $p$ or $q$ is odd.}
    \label{fig: 3PCF_Correlation}
\end{minipage}%
\hfill
\begin{minipage}[t]{.43\textwidth}
%\vspace{0pt}
  \centering
  \includegraphics[width=.9\textwidth]{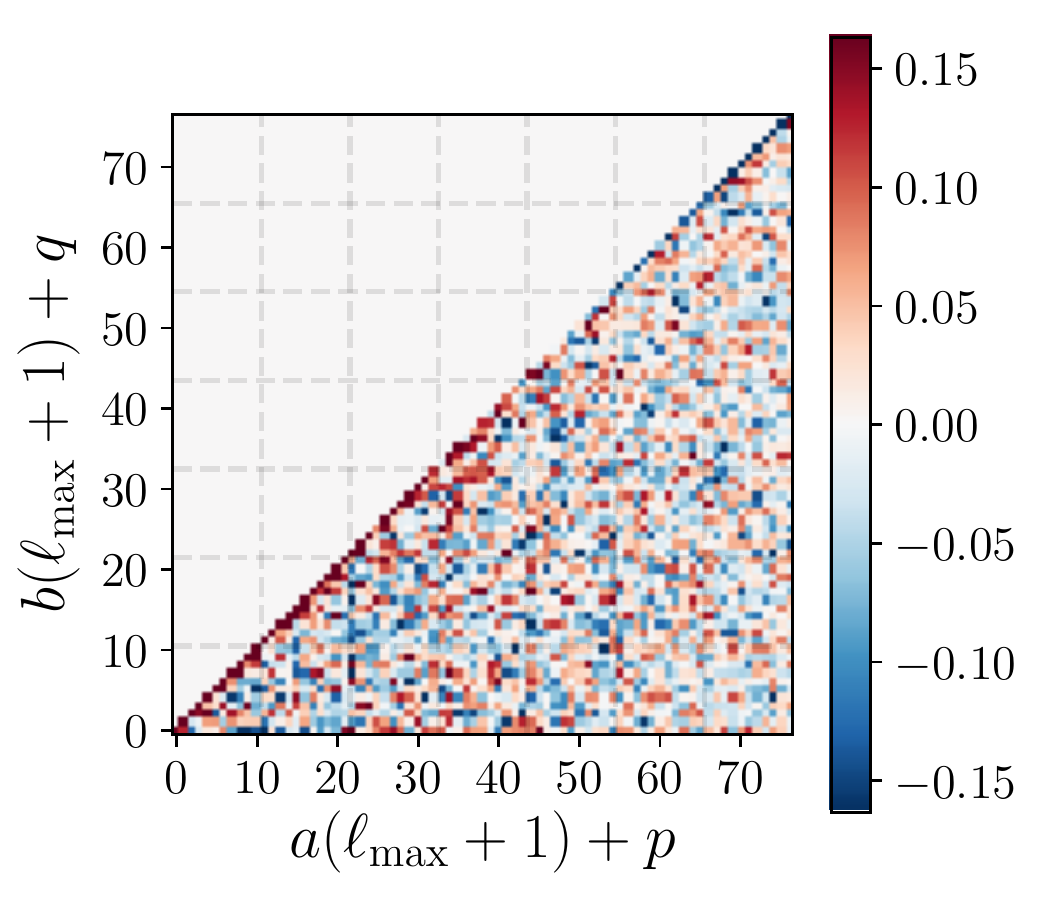}
  \caption{Discriminant matrix between sample and theoretical compressed 3PCF covariance matrices, as defined in Eq.\,\ref{eq: comparison_matrix}. The reduced theoretical matrix is shown in Fig.\,\ref{fig: 3PCF_Correlation} (in the same format as here) and the sample matrix is constructed from the covariance of 250 compressed 3PCF measurements, which are shown in Fig.\,\ref{fig: 3PCF-measurements}. As in Fig.\,\ref{fig: diff_matrix}, we mask the upper half of the matrix to avoid spurious correlations being observed from symmetry. If the theoretical matrix matches the data, we expect the discriminant matrix to be consistent with zero; as noted in the text, there is a slight excess of root-mean-square deviation here, concentrated along the leading diagonal. This indicates some small differences between sample and theoretical matrices, most likely due to the lack of proper inclusion of non-Gaussianity terms in our model.}
  \label{fig: 3PCF_Discriminant}
\end{minipage}
\end{figure}

\begin{figure}%
    \centering
    \subfloat[Mock Precision]{{\includegraphics[width=0.28\textwidth]{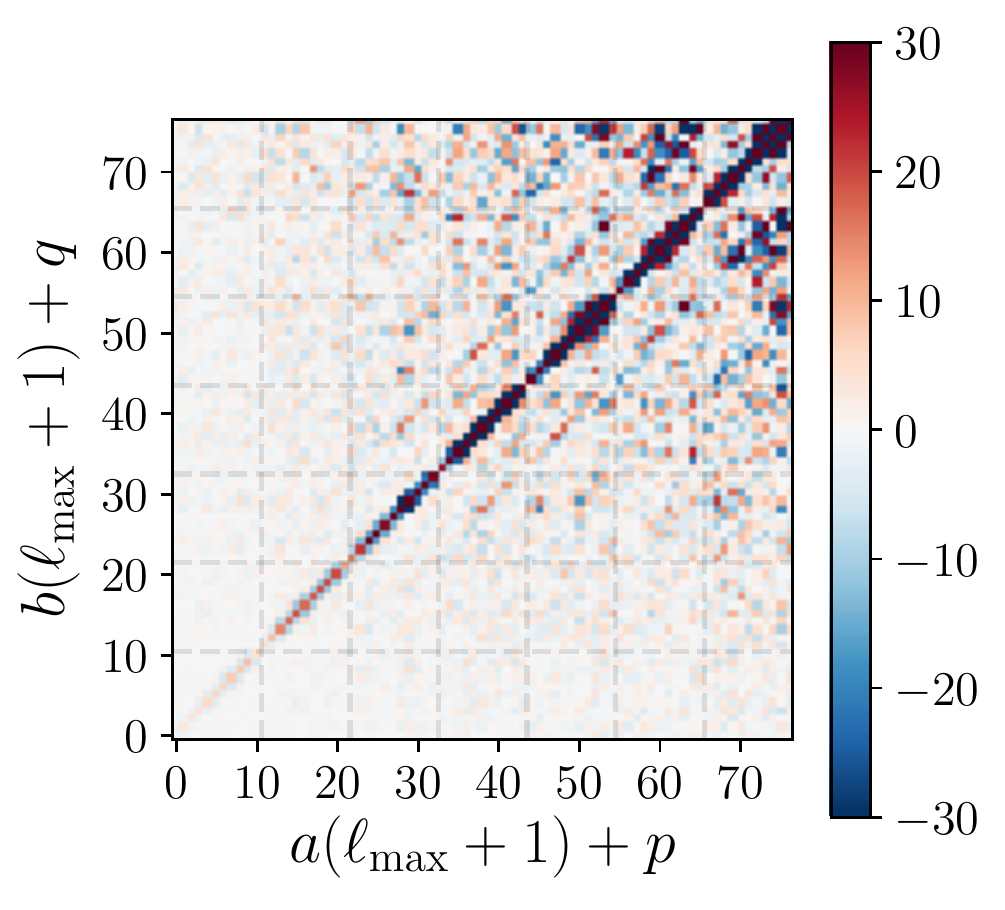} }}%
    \subfloat[Model Precision]{{\includegraphics[width=0.28\textwidth]{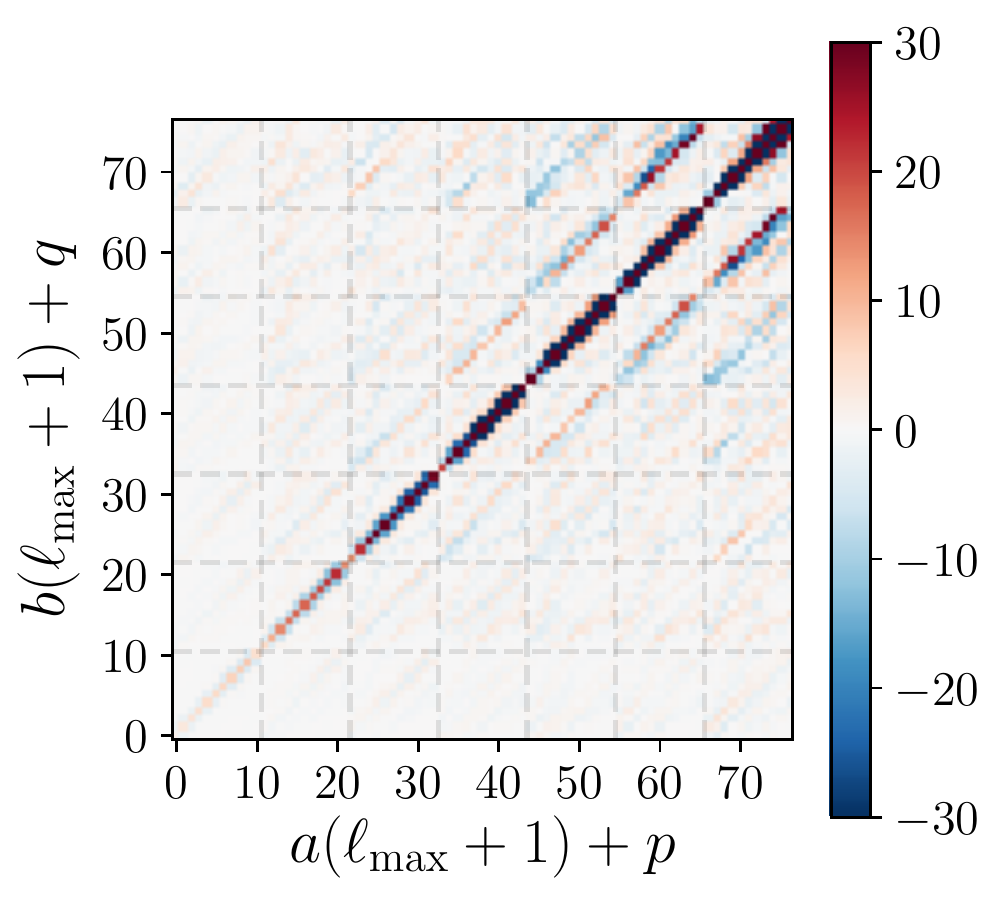}} }%
    \subfloat[(Mock - Model) Precision]{{\includegraphics[width=0.28\textwidth]{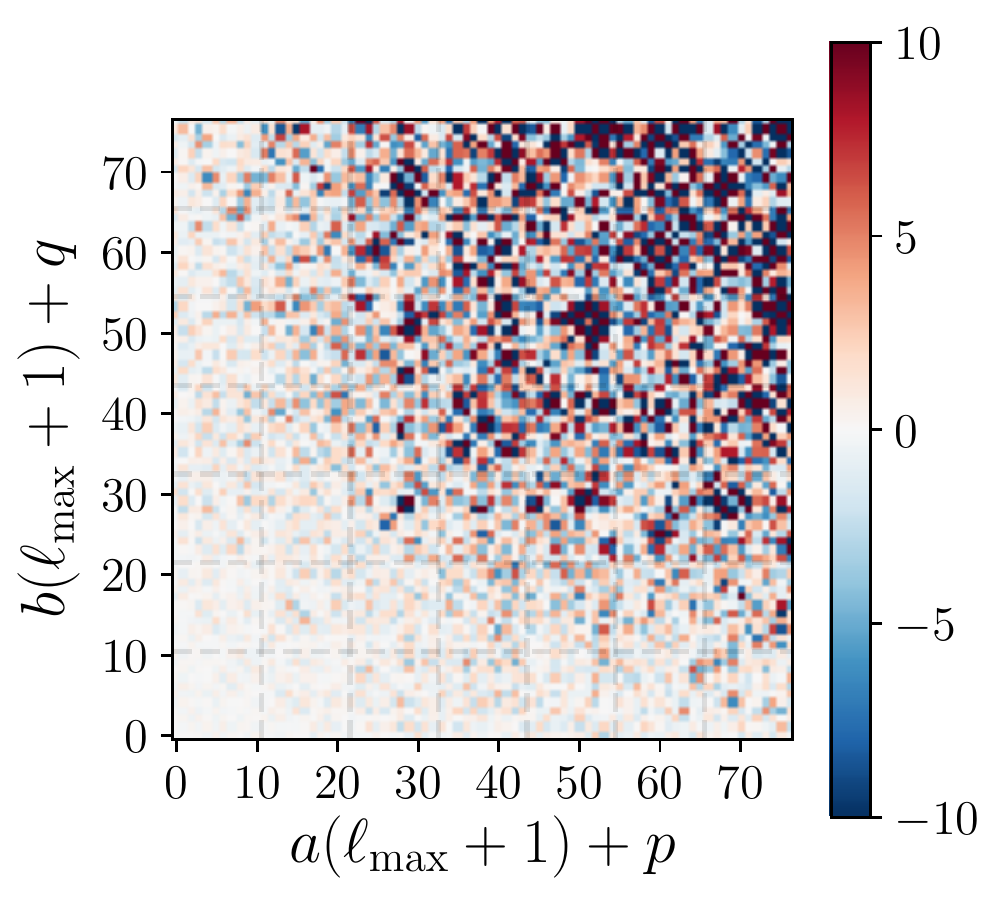}} }%
    \caption{Comparison of the sample and theoretical compressed 3PCF precision matrices across a number of radial bins and Legendre moments, in the same format as Figs.\,\ref{fig: 3PCF_Correlation}\,\&\,\ref{fig: 3PCF_Discriminant}, again using a theoretical shot-noise rescaling parameter of $\alpha = 1.20$. The sample (theoretical) compressed precision matrix is defined by Eq.\,\ref{eq: wishart-precision} (\ref{eq: non-wishart-precision}) and we remove the leading scalings by multiplying by $(2p+1)(2q+1)/(r_ar_b)^{2}$. We do not notice any systematic differences between the matrices except for larger noise at high multipole moment, as expected since the higher multipoles become significantly more noisy. An alternative comparison between the matrices is presented in Fig.\,\ref{fig: 3PCF_Discriminant}.}
    \label{fig: 3pcf-precisions}
\end{figure}

Given the set of 250 compressed 3PCF estimates $\left\{\hat{\zeta}^{(n),a}_{(c),p}\right\}$ with mean $\bar{\zeta}^a_{(c),p}$, the compressed covariance (denoted by $\mathbb{C}$) is defined in the standard manner;
\beq\label{eq: compressed-sample-covariance}
    \mathbb{C}^{a,b}_{p,q} = \frac{1}{N_\mathrm{mocks}-1}\sum_{n=1}^{N_\mathrm{mocks}}\left[\left(\hat{\zeta}^{(n),a}_{(c),p}-\bar{\zeta}^a_{(c),p}\right)\left(\hat{\zeta}^{(n),b}_{(c),q}-\bar{\zeta}^b_{(c),q}\right)\right],
\eeq
and is related to the uncompressed covariance by
\beq\label{eq: compressed-covariance}
    \mathbb{C}^{a,b}_{p,q} = \frac{\sum_{c\in S(a)}\sum_{d\in S(b)}C_{p,q}^{ac,bd}v_cv_d}{\sum_{c\in S(a)}\sum_{d\in S(b)}v_cv_d}
\eeq
where $C^{ac,bd}_{p,q} = \operatorname{cov}(\zeta^{ac}_p,\zeta^{bd}_q)$. The (symmetric) compressed matrix has a total of $\tfrac{1}{2}\left[N_r^{(c)}(\ell_\mathrm{max}+1)\right]\left[(N_r^{(c)}(\ell_\mathrm{max}+1)+1\right] = 3003$ independent components, where $N_r^{(c)}=11$ is the number of radial bins allowed by the compression strategy. In contrast, the uncompressed matrix has $\frac{1}{2}N_\mathrm{elem}(N_\mathrm{elem}+1)$ elements, for $N_\mathrm{elem}=\tfrac{1}{2}N_r(N_r+1)(\ell_\mathrm{max}+1)$, or greater than $10^6$ for $N_r = 15$. The large reduction in dimensionality provided by the compression thus significantly reduces matrix noise and gives much expedited integral convergence.

The theoretical compressed covariance is computed via the algorithm discussed in Sec.\,\ref{subsec: 3pcf-algorithm}, initially computing the fully symmetric matrix $C^{ab,cd}_{p,q}$ before applying the compression of Eq.\,\ref{eq: compressed-covariance}. The relevant three- to six-point integrals are sampled using positions taken from a $N_\mathrm{rand}=10N_\mathrm{gal}$ random catalog, with an additional input being a finely binned 2PCF estimated from the mean of 1000 QPM mocks, as in Sec.\,\ref{subsec: 2pcf-data-comparison}. To determine the three-point survey-correction function $\Phi(r_a,r_b,\chi)$, we first compute the $RRR$ triple counts of a small random catalog, using the output monopoles to define a smooth function (via Eq.\,\ref{eq: triple_continuous}), which is additionally normalized by $6V\overline{(nw)^3}$ for later convenience. For any given pair of bins, we found the inverse function $\Phi^{-1}(r_a,r_b,\chi)$ to be well fit by its first six multipoles, allowing for quick reconstruction when the code is run (noting that the relevant Legendre polynomials already must be computed by the sampler). In total, we sample $8.4\times 10^{12}$ hexes of particles (including those rejected for being outside the survey or in incorrect bins) over $260$ CPU-hours (with trivial parallelization possible). For a given number of hexes sampled, the computation time scales as $(\ell_\mathrm{max}+1)^2$, since we must add each accepted set to this many Legendre bins.

Since the width of the DR12 survey ($\sim 1600h^{-1}\mathrm{Mpc}$) is large compared to the characteristic scale of $\xi(\vec r)$ ($\sim 100h^{-1}\mathrm{Mpc}$) and the maximum bin-size ($180h^{-1}\mathrm{Mpc}$), we expect the first six-point covariance term to be strongly subdominant compared with the other terms (see Appendix \ref{sec: 3PCF-cancellation}). Computationally, this term poses difficulties, since cancellation occurs via negative contributions at large radius balancing positive ones at small $r$. Accurate computation of this term requires sampling of all possible values of the $k-l$ particle separation (cf.\,Fig.\,\ref{fig: sampling_diagrams}) which is inefficient, thus we here simply assume the term to be zero.

An important hyperparameter in our theoretical covariance matrix is the shot-noise rescaling parameter $\alpha$, used to approximate non-Gaussianity. In previous work \citep{2019MNRAS.487.2701O,rascalC} it was shown that, for the 2PCF, this could be constrained via splitting the survey into jackknife regions and comparing theoretical and jackknife covariance matrices. Whilst this remains possible for the 3PCF, we here adopt a simpler approach, computing $\alpha$ by fitting the theoretical to sample covariance matrix. Although this introduces dependence of our theoretical matrix on mock data, we note that we do not need a large number of mocks to constrain $\alpha$, since there is only a single degree of freedom. As before, $\alpha$ is determined by minimizing the Kullback-Leibler \citep{kullback1951} divergence between the matrices;
\beq
    \alpha^{*} &=& \operatorname*{arg\,min}_{\alpha}\left[D_{KL}(\hat{\mathbf{\Psi}}_\mathrm{theory}(\alpha),\hat{\mathbf{C}}_\mathrm{sample})\right]\\\nonumber
    D_{KL}(\mathbf{\Psi},\mathbf{C}) &=& \frac{1}{2}\left[\operatorname{trace}(\mathbf{\Psi}\mathbf{C}) - \log\det\mathbf{\Psi} - \log\det\mathbf{C} - N_\mathrm{bins}\right].
\eeq
We obtain an optimal value of $\alpha = 1.20$, significantly larger than that of the 2PCF ($1.032$) indicating that non-Gaussianity is more important in the 3PCF auto-covariance. Using $N_\mathrm{samples} = 40$ individual matrix estimates (each run on a different core) with this shot-noise rescaling, we obtain $N_\mathrm{eff}=2\times 10^4$ from Eq.\,\ref{eq: N_eff_computation}, implying that our co-added estimate has an precision comparable with that of $20,000$ mocks. Despite sampling the same number of particle sets as for the 2PCF auto-covariance in Sec.\,\ref{subsec: 2pcf-data-comparison}, we note a much reduced $N_\mathrm{eff}$; this is due to the higher dimensionality of our problem; sampling in 18-dimensions naturally gives slower convergence than 12-dimensions. We further note that $N_\mathrm{eff}$ scales linearly with the number of hexes sampled, thus greater precision (though not necessarily accuracy) can be obtained simply via longer integration time.

The resulting compressed 3PCF correlation matrix (defined analogously to the 2PCF case of Eq.\,\ref{eq: correlation_matrix}) is displayed in Fig.\,\ref{fig: 3PCF_Correlation}. Strongest correlations are seen between bins with the same multipole moment (i.e. $p=q$), though these are still non-negligible for bins with very different multipole moments. Noticeably, correlations involving odd multipoles are negative away from the leading diagonal and there is strong correlation of all bins with the quadrupole term, which is seen to be the dominant non-Gaussian component in Fig.\,\ref{fig: 3PCF-measurements}.  

A simple way to compare theory to simulation is via the precision matrices (computed via Eqs.\,\ref{eq: wishart-precision}\,\&\,\ref{eq: non-wishart-precision}), which are plotted in Fig.\,\ref{fig: 3pcf-precisions}. We choose to plot these rather than the covariance matrices since they have more relevance for parameter inference, though we require a large $N_\mathrm{mocks}$ to ensure invertibility (which is why the 2PCF sample precision was not shown in Sec.\,\ref{subsec: 2pcf-data-comparison}). Plotting the matrices multiplied by $(2p+1)(2q+1)/(r_ar_b)^2$ to remove the leading scalings, we note good agreement between mock and model precision for low multipoles, with significant random error in the mock data for higher $(p,q)$ pairs, due to the greater intrinsic noise in these 3PCF measurements. Indeed, plotting the difference of sample and theoretical covariance matrices (Fig.\,\ref{fig: 3pcf-precisions}c) shows no noticeable systematic differences between the two, although we are strongly noise limited at larger $(p,q)$. More mocks are required to further constrain systematic differences between the precision matrices. 

As in Sec.\,\ref{subsec: 2pcf-data-comparison}, we may additionally compare matrices via the discriminant matrix, $\mathbf{P}$, defined in Eq.\,\ref{eq: comparison_matrix} and displayed in Fig.\,\ref{fig: 3PCF_Discriminant}. Although the bulk of the matrix appears consistent with random fluctuations, there is a slight overestimate along the diagonal (where $a=b$ and $p=q$), which indicates some systematic differences between matrices. The mean and root-mean-square of $\mathbf{P}$ are here found to be $0.13\%$ and $5.8\%$ respectively, compared to the expected deviance of $4.9\%$ given the number of degrees of freedom and $N_\mathrm{mocks}$, which again indicates a slight bias. This difference is attributed to the lack of proper inclusion of non-Gaussian terms in our 3PCF covariance formalism and may be significantly reduced via simple models for the 3PCF, e.g. hierarchical models \citep{1975ApJ...196....1P,1977ApJ...217..385G}. However, the lack of any noticeable features in the precision matrix differential indicates that this may not have major impacts for cosmological analyses.

Holistically, it is clear that our 3PCF covariance matrix formalism is able to produce useful estimates for the survey-dependent 3PCF auto-covariance matrix which are in fair agreement with those from a suite of mocks, especially when considering the precision matrices. The success of our shot-noise rescaled estimates is perhaps surprising, given recent works such as \citet{2017PhRvD..96b3528C}, which claim that the non-Gaussian contributions to the 3PCF covariance are large even on mildly non-linear scales. We note however that (a) we require a larger shot-noise rescaling for the 3PCF than the 2PCF auto-covariance, indicating more non-Gaussianity, and (b) we have focused on scales above $40\,h^{-1}\mathrm{Mpc}$ (as appropriate for BAO analyses), where the effects of non-Gaussianity are expected to be weak. We expect our approximation to lose accuracy on small scales.

\section{Summary and Outlook}\label{sec: Conclusion}
In this paper we have developed a formalism for estimating theoretical anisotropic two-point and isotropic three-point correlation function covariance matrices in Legendre multipole bins for arbitrary survey geometries, extending the work of \citet{2016MNRAS.462.2681O}, \citet{2019MNRAS.487.2701O} and \citet{rascalC}. Our covariances have been derived in configuration space via shot-noise expansions and Wick contractions of the density field, giving a set of integrals from which the Gaussian covariance can be estimated. To incorporate Legendre multipoles, we introduced a model for the random pair or triple counts and an associated geometry-correction function, which allowed us to sample Legendre covariances directly, rather than constructing them from angularly binned covariances in post-processing. As in previous works, a shot-noise rescaling parameter was included to account for some non-Gaussianity; this may be calibrated either from a (small) set of mocks or jackknives and allows us to neglect higher order terms in the matrix expansion.

We have extended the previously used 2PCF algorithm to the new cases considered here, estimating the relevant integrals by sampling sets of two to six points in space, made highly efficient via importance sampling and careful ordering of terms. This has been incorporated into the publicly available \texttt{RascalC} code,\footnote{\href{https://RascalC.readthedocs.io}{RascalC.readthedocs.io}} giving 2PCF (3PCF) covariance matrices with negligible sampling noise in $\sim 10$ ($\sim 100$) CPU-hours for BOSS-like surveys. We demonstrated the accuracy of our estimates by comparison with sample covariance matrices derived from Quick Particle Mesh mocks appropriate for BOSS DR12 (which are computed at much larger computational expense). For the 2PCF we found excellent agreement between theory and simulations, with only a small bias found in the 3PCF matrices, though this was not found to have a noticeable impact on the precision matrices. We caution that we do not expect such good agreement between shot-noise rescaled theory and mocks on significantly smaller scales.

We expect that any differences between theoretical and sample matrices may be substantially reduced by including the 3PCF in our covariance integrals, either through some simple model, or as a fit to the measured data. A simple way to achieve this would be through hierarchical models \citep{1975ApJ...196....1P,1977ApJ...217..385G}, where the 3PCF is written as $\zeta(\vec r_1,\vec r_2) = Q\big[\xi(r_1)\xi(r_2)+\xi(r_1)\xi(|\vec r_1-\vec r_2|) + \xi(r_2)\xi(|\vec r_1-\vec r_2|)\big]$, for hierarchical parameter $Q$. Although it is well known that $Q$ varies as a function of triangle configuration \citep[e.g.][]{2003MNRAS.340..580T}, fitting for a constant $Q$ will allow for a more accurate inclusion of non-Gaussianity, though at the expense of significant additional computation time for the 3PCF matrices.

The isotropic 3PCF covariance integrals presented herein may be simply extended to the anisotropic case, which involves an additional two 3PCF parameters describing the orientation of a triangle with respect to the mean line of sight, allowing for proper inclusion of redshift space distortion effects. This will modify the integrals only by the insertion of additional orientation-dependent binning functions (or Legendre multipoles), but we note that both the anisotropic 3PCF and its covariance are difficult to visualize, due to their high dimensionality. In addition, without some suitable compression, the number of independent bins is large, leading to slow convergence of the output matrices. An additional extension is to the cross-covariance of the correlation functions with the power-spectrum; this will be discussed in future work.

The benefit provided by algorithms for approximate covariance matrix generation is many-fold; due to the approaches' smaller dependence on mocks, computational focus can be shifted towards creating a few high accuracy mocks, rather than a large number, simply to drive down sample covariance matrix noise. In addition, the vast speed boost allows for new avenues of exploration, such as full analyses of the dependence of covariances on the survey geometry and assumed cosmology, which are infeasible using conventional mock-based approaches. It is the hope of both authors that the approaches presented in this paper will be used in future analyses of survey data (especially those applying multiple statistics in concert) to compute high precision covariances at low computational effort, and therefore assist in placing strong constraints on cosmological parameters.

\section*{Acknowledgements}

We thank the referee for insightful comments helping to improve the clarity of this paper. OHEP acknowledges funding from the Herchel-Smith foundation. DJE is supported by U.S. Department of Energy grant DE-SC0013718 and as a Simons Foundation Investigator.

Some of the computations in this paper were run on the Odyssey cluster supported by the FAS Division of Science, Research Computing Group at Harvard University. Funding for SDSS-III has been provided by the Alfred P. Sloan Foundation, the Participating Institutions, the National Science Foundation, and the U.S. Department of Energy Office of Science. The SDSS-III web site is \href{http://www.sdss3.org}{www.sdss3.org/}.

SDSS-III is managed by the Astrophysical Research Consortium for the Participating Institutions of the SDSS-III Collaboration including the University of Arizona, the Brazilian Participation Group, Brookhaven National Laboratory, Carnegie Mellon University, University of Florida, the French Participation Group, the German Participation Group, Harvard University, the Instituto de Astrofisica de Canarias, the Michigan State/Notre Dame/JINA Participation Group, Johns Hopkins University, Lawrence Berkeley National Laboratory, Max Planck Institute for Astrophysics, Max Planck Institute for Extraterrestrial Physics, New Mexico State University, New York University, Ohio State University, Pennsylvania State University, University of Portsmouth, Princeton University, the Spanish Participation Group, University of Tokyo, University of Utah, Vanderbilt University, University of Virginia, University of Washington, and Yale University. 

\bibliographystyle{mnras}
\bibliography{full_lib}%adslib,otherlib} % if your bibtex file is called example.bib

%%%%%%%%%%%%%%%%%%%%%%%%%%%%%%%%%%%%%%%%%%%%%%%%%%

%%%%%%%%%%%%%%%%% APPENDICES %%%%%%%%%%%%%%%%%%%%%

\appendix

\section{Cancelling Term in the 3PCF Auto-Covariance}\label{sec: 3PCF-cancellation}
We here discuss the cancellation of the first Gaussian six-point term in the theoretical 3PCF auto-covariance expansion. From Eq.\,\ref{eq: cov3x3 full terms}, we may write the first Gaussian term in integral form as
\beq
    {}^6_A C^{ab,cd}_{p,q} &=& 9\frac{(2p+1)(2q+1)}{\left[6V\overline{(nw)^3}\right]^2v_av_bv_cv_d}\int d^3\vec r_i\,d^3\vec r_j\,d^3\vec r_k\,d^3\vec r_l\,d^3\vec r_m\,d^3\vec r_n\,n(\vec r_i)n(\vec r_j)n(\vec r_k)n(\vec r_l)n(\vec r_m)n(\vec r_n)\\\nonumber
    &\times&w(\vec r_i)w(\vec r_j)w(\vec r_k)w(\vec r_l)w(\vec r_m)w(\vec r_n)K^{ab,p}(\vec r_i,\vec r_j,\vec r_k)K^{cd,q}(\vec r_l,\vec r_m,\vec r_n)\xi(\vec r_i-\vec r_j)\xi(\vec r_k-\vec r_l)\xi(\vec r_m-\vec r_n)\\\nonumber
    &=& \int d^3\vec r_k\,d^3\vec r_l\,\mathcal{H}^{ab,p}(\vec r_k)\mathcal{H}^{cd,q}(\vec r_l)\xi(\vec r_k-\vec r_l)
\eeq
where we have separated out the integrals over $\vec r_k$ and $\vec r_l$ and defined the partial integral
\beq
    \mathcal{H}^{ab,p}(\vec r_j) &=& \frac{2p+1}{2V\overline{(nw)^3}v_av_b}\int d^3\vec r_i\,d^3\vec r_j\,n(\vec r_i)n(\vec r_j)w(\vec r_i)w(\vec r_j)K^{ab,p}(\vec r_i,\vec r_j,\vec r_k)\xi(\vec r_i-\vec r_j)\\\nonumber
    &=& \frac{2p+1}{2V\overline{(nw)^3}}\int d^3\vec r_i\,d^3\vec r_j\,n(\vec r_i)n(\vec r_j)w(\vec r_i)w(\vec r_j)\xi(\vec r_i-\vec r_j)\left[\Theta^a(\vec r_i-\vec r_j)\Theta^b(\vec r_i-\vec r_k)\Phi(r_a,r_b,\chi_{(\vec r_j-\vec r_k)})L_p(\chi_{(\vec r_j-\vec r_k)})+\text{5 perms.}\right]
\eeq
(expanding $K^{ab,p}$ in the second line) with an analogous form for $\mathcal{H}^{cd,q}(\vec r_k)$. Notably, $\mathcal{H}$ is simply a (weighted) integral of the 2PCF and a Legendre polynomial over all triangles with sides in bins $a,b$, with $\vec r_k$ specifying the position of one of the triangle vertices. For an ideal uniform survey with periodic boundary conditions, we would expect this integral to be invariant under spatial translations of the triangle, thus $\mathcal{H}(\vec r_k)$ should be independent of $\vec r_k$. In this limit, the covariance term may be simplified as
\beq
    {}^6_AC^{ab,cd}_{p,q} = \mathcal{H}^{ab,p}\mathcal{H}^{cd,q}\int d^3\vec r_k\,d^3\vec r_l\,\xi(\vec r_k-\vec r_l) = \mathcal{H}^{ab,p}\mathcal{H}^{cd,q}\times V\int d^3\vec r_{kl}\,\xi(\vec r_{kl})
\eeq
transforming $\vec r_k$ to $\vec r_{kl} \equiv \vec r_k-\vec r_l$ and integrating over the survey volume $V$. Noting that the 2PCF is defined as an over-random probability, we expect its infinite integral to be zero, such that the large-scale mean galaxy number density is equal to the density in the absence of correlations. In practice, we are limited by the finite volume of the survey and the lack of knowledge of the true uncorrelated number density, but the 2PCF integral (equal to the zero-momentum power spectrum $P(\vec 0)$) is expected to be small. 

For a realistic survey, we will expect some dependence of $\mathcal{H}$ on position (due to boundaries and non-uniform number densities and weights) but, assuming the characteristic survey size $L$ is much larger than the maximum bin-width, this dependence is expected to be weak. Furthermore, if $L$ is large compared to the characteristic scale of $\xi(\vec r)$ (i.e. the BAO scale $\sim 100h^{-1}\mathrm{Mpc}$) the 2PCF integral will be small, thus the entire term is expected to be negligible compared to the other six-point covariance integral, for all combinations of bins and Legendre moments.

%%%%%%%%%%%%%%%%%%%%%%%%%%%%%%%%%%%%%%%%%%%%%%%%%%

% Don't change these lines
\bsp	% typesetting comment
\label{lastpage}
\end{document}